\documentclass[final,3p,times,twocolumn]{elsarticle}

\usepackage{amssymb}

\newcommand{\be}{\begin{equation}}
\newcommand{\ee}{\end{equation}}
\newcommand{\ba}{\begin{eqnarray}}
\newcommand{\ea}{\end{eqnarray}}
\newcommand{\nn}{\nonumber\\}
\def\gs{\mathrel{\raise1.16pt\hbox{$>$}\kern-7.0pt %
\lower3.06pt\hbox{{$\scriptstyle \sim$}}}}         %
\def\ls{\mathrel{\raise1.16pt\hbox{$<$}\kern-7.0pt %
\lower3.06pt\hbox{{$\scriptstyle \sim$}}}}         %

\journal{New Astronomy Reviews}

\begin{document}

\begin{frontmatter}

\title{Image Analysis for Cosmology: \\ Shape Measurement Challenge
  Review \& \\ Results from the Mapping Dark Matter Challenge}


\author[label1]{T. D. Kitching}
\ead{tdk@roe.ac.uk}
\author[label2,label3]{J. Rhodes}
\author[label1]{C. Heymans}
\author[label4]{R. Massey}
\author[label13]{Q. Liu}
\author[label5]{M. Cobzarenco}
\author[label6]{B. L. Cragin}
\author[label7]{A. Hassa\"{i}ne}
\author[label8]{D. Kirkby}
\author[label9]{E. Jin Lok}
\author[label8]{D. Margala}
\author[label10]{J. Moser}
\author[label11,label15]{M. O'Leary} 
\author[label14]{A. M. Pires}
\author[label12]{S. Yurgenson}

\address[label1]{SUPA, Institute for Astronomy, University of Edinburgh, EH9 1RZ, UK}
\address[label2]{Jet Propulsion Laboratory, California Institute of
  Technology, 4800 Oak Grove Drive, Pasadena, CA 91109, USA}
\address[label3]{California Institute of Technology, 1200 East California
Boulevard, Pasadena, CA 91106, USA}
\address[label4]{Institute for Computational Cosmology, Durham University, South
Road, Durham, DH1 3LE, U.K.}
\address[label13]{Physics Department, Columbia University, 538 West
  120th Street, 704 Pupin Hall, MC 5255 New York, NY 10027, USA}
\address[label5]{Department of Computer Science, University College London, Gower Street, London, WC1E 6BT, U.K}
\address[label6]{Department of Physics, Keene State College, Keene, NH
  03435, USA}
\address[label7]{Computer Science and Engineering Department, College
  of Engineering, Qatar University, P.O.Box 2713, Doha, Qatar}
\address[label8]{Department of Physics and Astronomy, UC Irvine, 4129 Frederick
Reines Hall, Irvine, CA 92697-4575, USA}
\address[label9]{Deloitte Analytics, Level 18, 550 Bourke Street, Melbourne, VIC, 3000, Australia}
\address[label10]{Kaggle, 665 Third Street, San Francisco, CA 94107, USA}
\address[label11]{Department of Atmospheric, Oceanic and Space Sciences, University of Michigan, 2455 Hayward St, Ann Arbor, MI 48109, USA}
\address[label15]{Scott Polar Research Institute, University of Cambridge, Lensfield Road, Cambridge, CB2 1ER, UK}
\address[label14]{Department of Mathematics and CEMAT, IST, TULisbon, Av.
Rovisco Pais, 1049-001 Lisboa, Portugal}
\address[label12]{Harvard Medical School, 25 Shattuck Street Boston, MA
  02115., USA}

\begin{abstract}
In this paper we present results from the Mapping Dark
Matter competition that expressed the weak lensing shape measurement task in its
simplest form and as a result attracted over 700 submissions in 2
months and a factor of 3 improvement in shape measurement accuracy on
high signal to noise galaxies, over previously published results, and a
factor 10 improvement over methods tested on constant shear blind simulations. 
We also review weak lensing shape measurement challenges, 
including the Shear TEsting Programmes (STEP1 and
STEP2) and the GRavitational lEnsing Accuracy Testing competitions
(GREAT08 and GREAT10). 
\end{abstract}

\begin{keyword}
Cosmology \sep Image Analysis \sep Gravitational Lensing \sep Dark
Energy \sep Dark Matter 
\end{keyword}

\end{frontmatter}

% \linenumbers

\section{Introduction}
\label{Intro}
Image analysis in cosmology is a process that involves  
taking pixelised and noisy images of objects, extracting information
from them, and using these to infer properties of the large scale structure of the Universe. 
This is of paramount importance for the  
endeavour of understanding dark matter and dark energy, those
phenomena whose mass-energy account for approximately 26\% and 70\%
of the Universe respectively  
and whose fundamental nature is entirely unknown. Of particular interest is \emph{weak lensing} 
that has
been identified as one of the primary tools with which we can  
map the large scale structure and evolution of the Universe
(see reviews e.g. Albrecht et al., 2006; Peacock et al., 2006; Massey,
Kitching, Richard, 2010; Bartelmann \& Schneider, 2001; Weinberg et
al., 2012 and references therein).

\begin{figure*}
  \vspace{-3.0cm}
  \centerline{\includegraphics[width=1.5\columnwidth,angle=90,clip=]{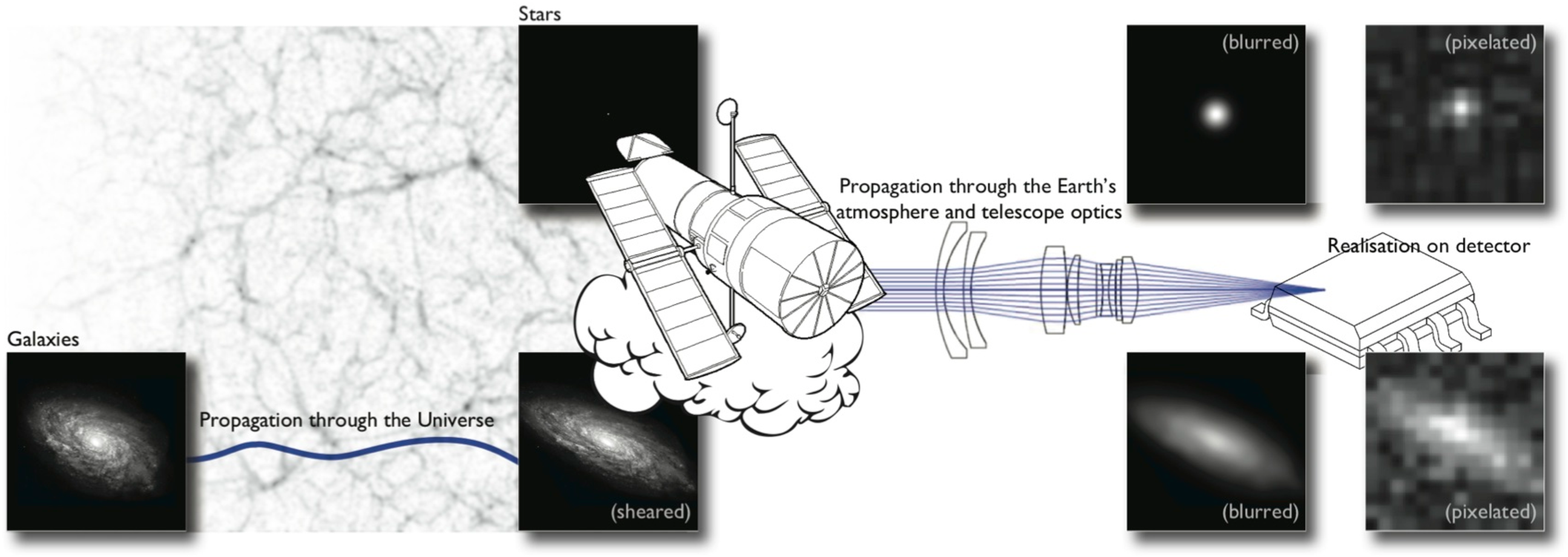}}
  \vspace{-3.0cm}
 \caption{This figure is reproduced from the GREAT10 Handbook
   (Kitching et al., 2011) with permission. As light propagates through the large
   scale structure of the Universe an additional ellipticity `shear'
   is imprinted on a galaxy's observed image. We observe sheared galaxies in the
   presence of a blurring convolution kernel (PSF), pixelisation from detectors and in the
   presence of noise. Shape measurement algorithms must be designed that
   measure the ellipticity of galaxies in the presence of these effects
   to enable the statistical properties of the shear field to be
   inferred. Star images can be used to estimate the PSF, since they
   approximate a point-source response to the convolution and pixelisation but are not affected by
   the shear.}
 \label{forward}
\end{figure*}

Weak lensing is the effect whereby the integrated
mass along the line of sight acts to induce an additional ellipticity  
to the observed light profile of an object, this additional
ellipticity is called shear. Distant galaxies have a measurable
additional ellipticity, because of the  
large amount of integrated mass along the line of sight, but local
objects do not. If we can therefore measure the ellipticity of distant
galaxies we can  
make statistical statements about the properties of the intervening
distribution of matter; see Figure \ref{forward}. 
These statements are necessarily statistical because for an individual  
object the additional ellipticity cannot be disentangled from
the object's `intrinsic' (un-sheared) ellipticity; and to make matters  
worse galaxies are inherently elliptical. However we can assume
that on average there is no preferred orientation for galaxies in the
Universe, that the mean
ellipticity should be zero if there were no intervening mass.   
Therefore by averaging over many galaxies any residual shear can then be
attributed to the matter distribution. In general cosmological
information comes not from the mean but the variance of the  
ellipticities (see Kitching et al., 2011).   

In fact there are two `modes' of using weak lensing data to
investigate the dark matter
distribution, both are statistical but treat the data and observations
in different ways. One is a `holistic' measure (we use the word in its meaning
of emphasising the importance of the whole and the interdependence of
its parts) where power spectra/correlation functions are created: 
one averages over all galaxies in a survey
and determines the two-point (or more generally n-point) functions and
compares these to theoretical predictions. The second approach is
`atomistic' where we also look at individual mass peaks and make dark
matter maps: one identifies individual objects of interest
(e.g. galaxy clusters) and generates a visual map of dark matter. 
 
The task of measuring the weak lensing effect is particularly difficult
because of noise in the images, pixelisation, and that we do not know  
in detail how to model the surface brightness distribution of
undistorted galaxies. As a result of these difficulties  
many methods have been proposed to measure the weak lensing effect,
either using direct model-independent pixel-level extraction of
parameters (for example Kaiser, Squires \& Broadhurst, 1995; Melchior et al., 2011) 
or using forward modelling of the galaxies (for example Kuijken, 1999;
Refregier 2003; Miller et al., 2007; Kitching et al., 2008). 

Importantly for weak lensing, to test the ability of a method to
extract the shear information from an ensemble of galaxies we cannot
take an observation  
that removes the shear effect, and because of the statistical nature
of the shear information we cannot compare the fidelity of an
individual object's  
inferred shear against what we would have hoped to observe in the
presence of perfect data. This is in contrast to photometric redshifts
for example where a spectra  
of an individual object can be taken and compared to the
photometrically inferred redshift estimate. To test shape measurement
methods we therefore must have  
accurate simulations whose aim is to test fidelity of these methods under
controlled conditions. 

Within the weak lensing community a number of
such simulations were started  
and run as competitions/challenges (the Shear TEsting Programme, STEP;
Heymans et al. 2006, Massey et al. 2007) under blind conditions,
which are a necessity so that algorithms cannot be tuned with
calibration factors. Reaching  
beyond the weak lensing community these competitions were opened up
to public participation (the GRavitational lEnsing Accuracy Testing,
GREAT08 and GREAT10; Bridle et al., 2009, Kitching et al., 2012) in an effort to spawn
new ideas and approaches to this 
algorithmic challenge. In this article we will review previous shape
measurement challenges, we will also present results from the most
widely participated and successful of these to date, the
Kaggle\footnote{{\tt http://www.kaggle.com/c/mdm}} Mapping Dark Matter
challenge,  
which attracted over 700 submissions in two months and saw an
improvement in the achieved accuracy of shape measurement methods by a
factor $3$, over previously published results (Bernstein, 2010 and
Gruen et al., 2010), and a
factor $10$ improvement over methods tested on blind simulations. 

This article is arranged as follows in Section 
\ref{Shape measurement challenges} we will review shape measurement  
challenges STEP and GREAT, and we
refer the reader to Kitching et al., (2011, 2012) for a full review of the
GREAT10 challenge. In  
Section \ref{Mapping Dark Matter} we will present the Mapping Dark Matter challenge simulations
and results as well as some commentary on the nature of setting
crowdsourcing challenges in astronomy. In Section \ref{Conclusions} we will discuss conclusions. 

\section{Shape measurement challenges}
\label{Shape measurement challenges}
Because we can never observe the unlensed ellipticity of objects algorithms that attempt to measure
shear parameters must be tested against simulations. In these
simulations a set of simulated galaxies are 
sheared by a known amount and this ‘true/simulated shear’ is compared to the measured shear
provided by the algorithms.

There are five publicly available lensing simulations from three related programmes: STEP (the
Shear TEsting Programme), GREAT (the GRavitational lEnsing Accuracy
Testing) and Mapping Dark Matter (more information
can be found here {\tt http://www.greatchallenges.info}). We summarise the main features
of these simulations in Table \ref{sims}. In the following we describe
the challenges STEP1, STEP2, GREAT08 and GREAT10 to provide context
for the Mapping Dark Matter results in Section \ref{Mapping Dark
  Matter}; these descriptions are pedagogical and describe the broad
motivation behind each of the simulation efforts. 
\begin{table*}\footnotesize
\centering
\begin{tabular}{|l|c|c|c|c|c|}
\hline
&{\bf STEP1} & {\bf STEP2} & {\bf GREAT08} & {\bf GREAT10} & {\bf MDM}\\
\hline
Galaxy Model&Simple&Complex(shapelets)&Simple&Simple(non-coelliptical)&Simple(non-coelliptical)\\
PSF Model&Simple(w/diff. spikes)&Realistic(ground)&Simple(Moffat)&Simple(Moffat)&Simple(Moffat)\\
PSF Knowledge&Unknown&Unknown&Known(functions)&Known(functions)&Known(pixelated images)\\
PSF Variation&Constant(unknown)&Constant(known)&Constant(known)&Variable(known)&Variable(known)\\
Object
Positions&Random(unknown)&Random(unknown)&Gridded(known)&Gridded(known)&Postage
Stamps(known)\\
Shear Variation&Constant&Constant&Constant&Variable&Constant\\
\hline
$N_{\rm galaxies}$&$\sim 0.7$x$10^6$&$\sim 2$x$10^6$&$30$x$10^6$&$50$x$10^6$&$0.1$x$10^6$\\
Metrics&$m$, $c$, $q$&$m$, $c$&$m$, $c$, $Q_{08}$&$m$, $c$, $q$,
$Q_{10}$, $\alpha$, $\beta$, ${\mathcal
  M}$, ${\mathcal A}$&RMSE, $m$, $c$, $Q_{08}$\\
Publicity&Shear Community&Shear Community&Open&Open&Open\\
Teams(Subs)&14&16&9(50)&9(100)&73(760)\\
Reference&Heymans et al. 2006&Massey et al. 2007&Bridle et
al. 2010&Kitching et al. 2012&this article\\
\hline
\end{tabular}
\caption{A summary of the main features of each shape measurement
  challenge to date (c. 2012), the metrics used in the analysis and
  some details of the accessibility of the challenge. $N_{\rm
    galaxies}$ is the approximate number of galaxies in the
  simulations. The number of teams is shown and the number of
  submissions in brackets.} 
\label{sims}
\end{table*}

\subsection{STEP1}
STEP1 was run in 2005 as the first programme in which shear
simulations were generated and tested by shape measurement methods
under blind conditions.  It was
inspired by the fact that there had been at least nine attempts to
measure the amplitude of the variance of matter fluctuations on 8 Mpc
scales, $\sigma_8$, from different data sets using different shape
measurement methods and it was found that these measurements disagreed at
the $2$-$\sigma$ level. It was suspected that shape measurement methods
may be the source of this discrepancy and it was decided that methods should
be tested in a blind way. 

The motivation behind this first challenge was to generate 
realistic astronomical images, using existing image generation software at
the time, and ask the question: 
\begin{center}
\emph{Can existing pipelines (including
source detection, PSF estimation and shape measurement) measure shear
accurately enough for current (c. 2006) data?}
\end{center}
The software used was {\tt SkyMaker}\footnote{\tt 
http://www.astromatic.net/software/skymaker}. The
images contained simulated galaxies and stars distributed in a
realistic manner across the images. The galaxies had models that
contained bulge plus disk components. There were six separate types
of PSF that were constant across the images, the PSFs had models that
ranged from circular Moffat functions to more complex functions that included diffraction
spikes. Participants were not told the PSF, or whether it was constant
or varying across the field of view, but asked to estimate it as
they would in real data. 

For each of the five PSF types there were 5 different values of the
shear (constant across the images) with
$\gamma_1=(0.0,0.005,0.01,0.05,0.1)$ and $\gamma_2=0.0$. This meant 
for each different PSF type, and $5$ different shear values, there
were $30$ different data sets, and each set consisted of $64$ different
images. Participants were asked to measure the shear in each image, there were
no rules on which galaxies should be used or how the shear was
estimated, and indeed participants were not even told how many
galaxies there were or whether objects were stars or galaxies. The
challenge then was to test the entire pipeline from source detection
and identification through to PSF estimation and shape measurement, 
in this respect the simulations were relatively
realistic and well matched to the question posed. The submitted shear
values, that were kept constant in each image were scored using a
metric that related the true input shear to the measured shear values 
\be
\gamma^M_i=(1+m_i)\gamma^T_i+c_i + q\gamma_i^2
\ee
for each shear component $i$, with a `multiplicative bias' $m$ and a
`constant bias' $c$; a perfect method would achieve results consistent
with $m=0$ and $c=0$. The quadratic term differs from that used
subsequently in GREAT10 that used $q\gamma_i|\gamma_i|$. 

\begin{figure*}
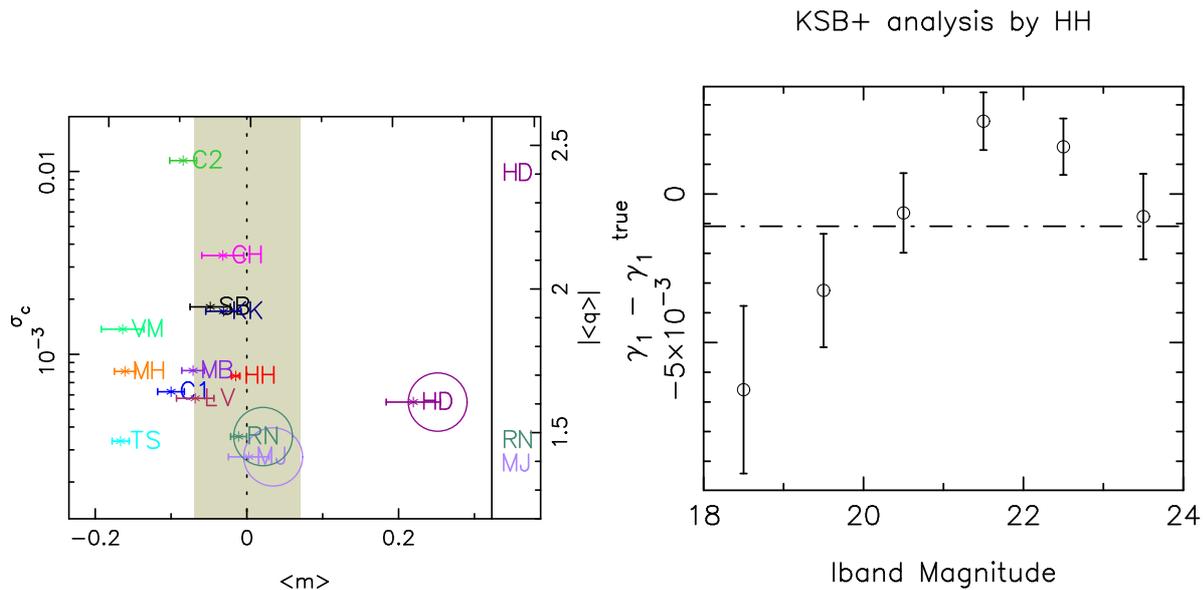

  \includegraphics[width=\columnwidth,clip=]{mbar_sigc_qbar.ps}
  \includegraphics[width=\columnwidth,clip=]{HH_mag_example.ps}
  \caption{These figures are reproduced from the STEP1 results
    (Heymans et al., 2006) with permission. The left panel shows the multiplicative
    bias $m$ against the variance on the constant bias $c$, methods
    that had a strong non-linear behaviour were circled and their $q$
    values shown. The right hand panel shows an example of how a
    particular methods true minus measured shear (`KSB+ HH', an implementation of Kaiser,
    Squires \& Broadhurst, 1995) varied as a function of simulated
    i-band magnitude.}
 \label{s1}
\end{figure*}

The STEP1 results (see Figure \ref{s1} for a selection) demonstrated
that the methods that were available at the time achieved an
accuracy that was sufficient for data sets available at that time. However there
was evidence for strong selection effects, biases that changed
depending on whether participants made false detections of objects, and
some strong condition-dependent biases (for example biases that varied
in a non-obvious way as a function of magnitude).  

\subsection{STEP2}
STEP2 was the second in the series of community challenges and was
launched soon after STEP1. Encouraged by the results of STEP1 the next
`step' was to complexify the simulations to lend further
credence to the existing methods abilities to measure shear for data
that existed at that time (c. 2007). The key area that was identified
as being not realistic in STEP1 was that the galaxy models used were
simple sums of exponential Sersic functions. At the same time a shape
measurement method `shapelets' (Refregier, 2003; Massey \& 
Refregier, 2005) was developed that made use of
sums of 2D basis functions to model complex galaxy morphologies, it
was realised that this approach could also be used to generate
simulations where each galaxy was constructed using shapelets. 
This enabled galaxies to be simulated with spiral aims, star forming regions
and simulated merging and irregular galaxies, using the image simulation
code {\tt SImage} (Massey et al., 2005; Ferry et al., 2008; Dobke et
al, 2010). 

As a further sophistication it was realised that ``shape noise'' was
a potentially dominating factor in shape measurement accuracy determination, where
the variance of the intrinsic (unsheared) ellipticities of galaxies
meant that a large number of simulations were needed to reduce this
term though Poisson statistics. To circumvent this issue it was
realised that if galaxies were simulated in pairs which had the same
shear but intrinsic ellipticities with opposite signs then when
averaging the observed ellipticity over the pair the intrinsic
ellipticity contribution would cancel to first order. This is captured
in the following average over such a pair
\be
\hat\gamma=[(e^{\rm int}+\gamma)_{\rm unrotated}+(-e^{\rm
    int}+\gamma)_{\rm rotated}]/2
\ee
where $\hat\gamma$ is the estimated shear, $e^{\rm int}$ the unsheared
intrinsic ellipticity and $\gamma$ the true shear, we show only first
order terms. This transform $e\rightarrow -e$ corresponds to a 90
degree rotation in the source image plane. 
This meant that images came in pairs one rotated by 90 degrees before
the shear was added and the other
unrotated, but participants were not aware which of the images was the
corresponding partner.
 
STEP2 had a similar simulation structure to STEP1, there were 6 
different PSF types, each was constant across the field of view and
had a complex profile, in particular for STEP2 the PSFs were simulated
by measuring the PSF using the shapelet decomposition from
a real ground-based telescope Subaru. The 90 degree rotated pairs
meant that the number of images needed per shear value was much
smaller (the 64 images per shear value in 
STEP1 were required to remove shape noise) so that
only 2 images (1 rotated/un-rotated pair) were required per shear
value. This meant that by keeping the simulation size approximately the
same many more shear values could be investigated, which meant the
simulation could not be reverse-engineered. STEP2 contained 128
images per PSF which meant 64 shear values per PSF and $128$x$6$ images in
total. All other realistic effects from STEP1 were kept, except that 
participants did not have to identify stars from galaxies in the images. 
The metric used to evaluate methods in STEP2 was again the $m$ and $c$ 
parameters defined for STEP1. 

\begin{figure*}
  \includegraphics[width=\columnwidth,clip=]{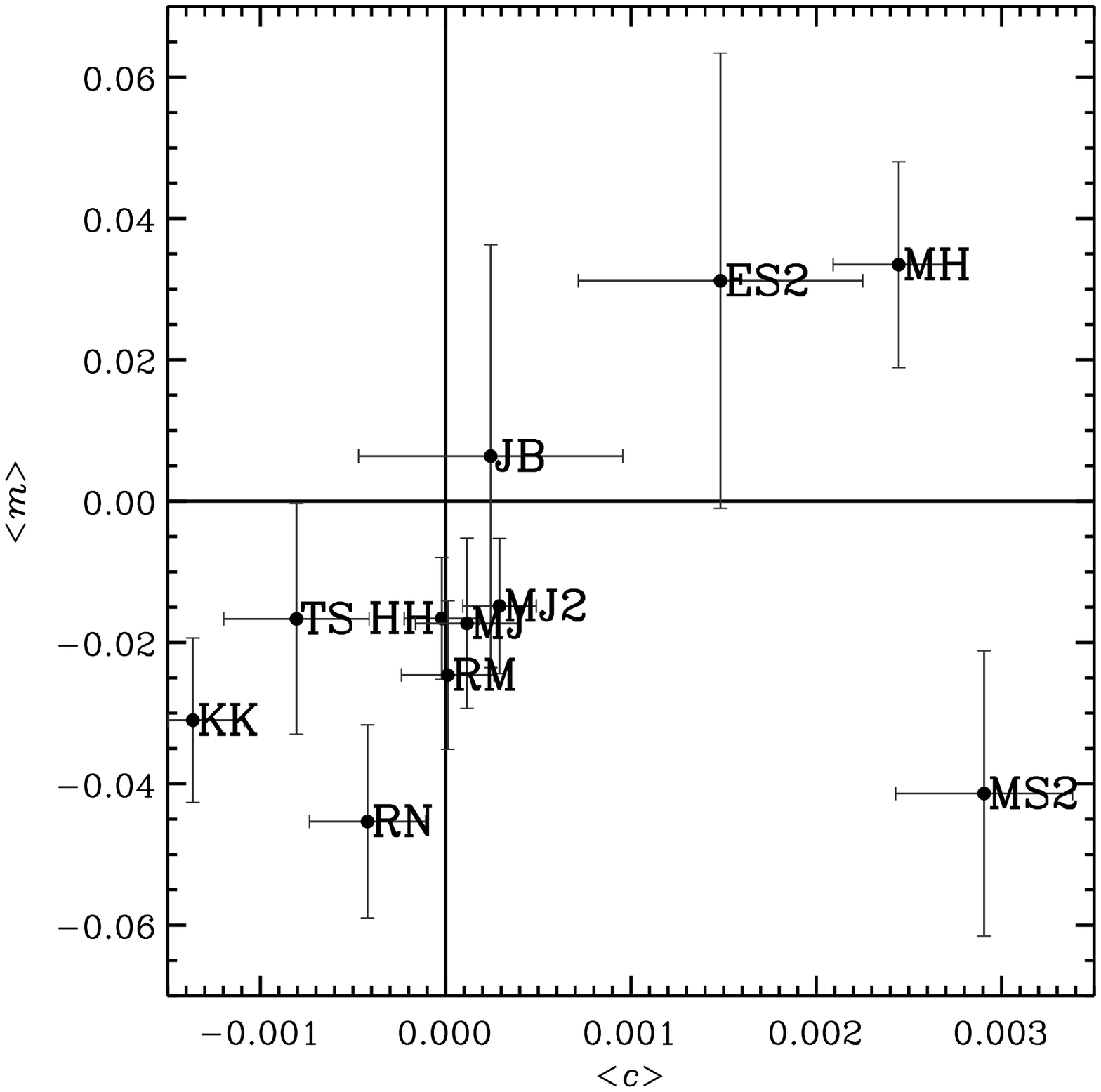}
  \includegraphics[width=\columnwidth,clip=]{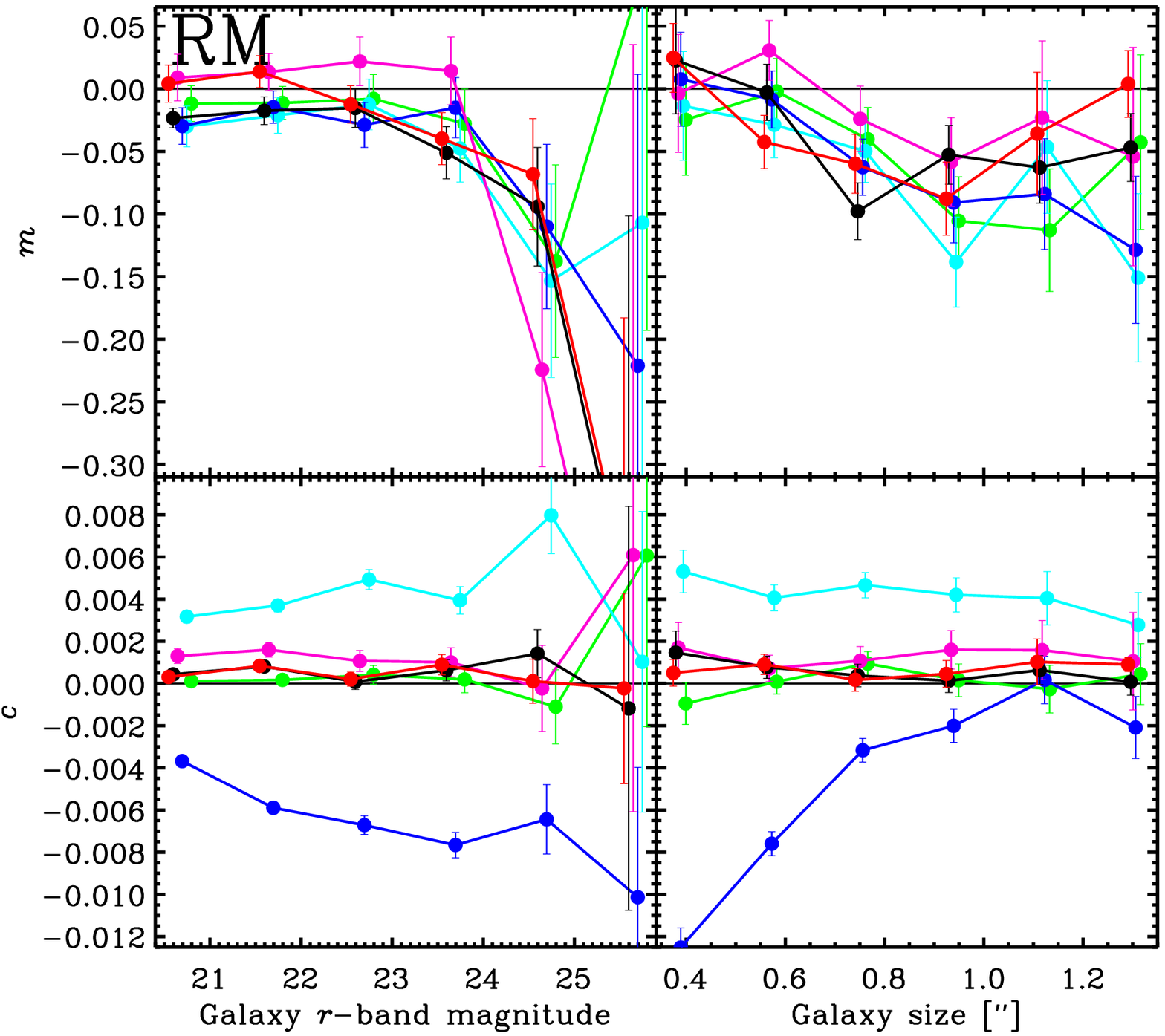}
  \caption{These figures are reproduced from the STEP2 results
    (Massey et al., 2007) with permission. The left panel shows the $m$ and $c$ values
    for each method that participated in STEP2. The right hand panel shows an example of how a
    particular methods (`RM', an implementation of shapelets Massey
    \& Refregier, 2005) $m$ and $c$ values varied as a function of simulated
    r-band magnitude and galaxy size.}
 \label{st2}
\end{figure*}

The STEP2 results (see Figure \ref{st2} for a selection) again
demonstrated that for the data available at the time (c. 2007) the
shape measurement methods available were sufficient. 
Similarly to STEP1 however there was no method that performed well as
a function of galaxy magnitude and size.

\subsection{GREAT08}
In the conclusion of STEP2 it was not clear what
aspect of the shape measurement methods were causing the biases in
particular regimes. There was also a shift in focus in the community
from an emphasis on parameters
such as $\sigma_8$ towards dark energy parameters 
as it was becoming clear that weak lensing is a
particularly good way of determining dark energy properties. Several 
authoritative reports were published in late 2006 highlighting this
fact (Albrecht et al., 2006; Peacock et al., 2006) such that by late
2007, when the STEP2 results were being scrutinised there was an 
new imperative for weak lensing studies. These realisations, with the fact 
that shape measurement biases were not understood in
detail, added a new impetuous to the task
of shape measurement. GRavitational lEnsing Accuracy
Testing 2008 (GREAT08) was then conceived where the aim was to reduce the
problem to its simplest expression (however in fact there were simpler
expressions found subsequently, see Section \ref{Mapping Dark Matter})
in order to determine if in the simplest case shape measurement
could work and to determine how and why shape measurement methods
biases were arising. 

An additional motivation was a further realisation
that in fact the problem is not an `astronomical/cosmological' problem
but an image analysis problem that could be accessible to
non-cosmologists, in particular computer scientists. In this tradition
the simulations were run as a competition (sponsored by
PASCAL\footnote{{\tt http://www.pascal-network.org/}}) with 
`winners' that were awarded prizes. The questions posed by GREAT08:
\begin{center}
\emph{Can we measure shapes under ideal circumstances? Why and how are
shape measurement methods biased?}
\end{center}
were qualitatively different to that posed by STEP, that focussed on
the direct usefulness of methods on simulations that were as
realistic as possible.   

The key changes from STEP2 were to provide participants with an
exact prescription for the PSF, as a functional form, to arrange galaxies
on a grid with known position and known type; source detection and
identification were not part of the challenge. The challenge again
used constant shear values across an image and the rotated-unrotated
method for reducing the simulation size. In order to encourage
participation GREAT08 used a live leaderboard where, instead of
methods submitting to the organiser (as in STEP1 and STEP2), the
submissions were uploaded to a server that automatically computed a
score. For this challenge a new metric was created that was the
inverse of the mean square error of the true and measured shear
\be 
Q_{08}\equiv\frac{10^{-4}}{\langle(\langle g^m_{ij}-g^t_{ij}\rangle_{j
    \in k})^2\rangle_{ik}}
\ee
where the averages were over the shear components and the images in
the challenge. This relates to the STEP $m$ and $c$ in a simple way,
but does not capture all useful information, the metric is mostly
sensitive to $c$, and is dependent on any
noise present in a method (see Kitching et al., 2008). This metric
however does provide a measure for a methods performance and meant
that the leaderboard feedback could not be reverse-engineered to
trivially calibrate methods in order to win the challenge. The
numerator was defined such that methods tested on STEP1 and STEP2
would have $Q_{08}\ls 50$ (see Figure \ref{qm3}) and methods that were limited only
statistically (by pixel-noise in the size of the simulated data set)
would achieve $Q_{08}\simeq 1000$.

GREAT08 was a success in its goals to attract non-cosmologists to the
problem in that the winner, and 2 out of 9 teams, 
were computer scientists. Methods used
previously in STEP performed at approximately the same level. A number of
clear trends were identified including that methods were biased in
particular at low signal to noise and for small galaxies (relative to
the PSF size). The best performing methods used the fact that the
shear was constant across each image to stack all galaxies together
(either in real or Fourier space) to cancel out intrinsic ellipticity
further and any noise, such that it was not clear how these methods, whilst performing 
well in this regime, were applicable to real data.   

\subsection{GREAT10}
The conclusion of STEP1, STEP2 and GREAT08, designed to test methods using constant
shear simulations was that the best method to do this was to stack all
the galaxies in the images. Unfortunately such a method, stacking all
galaxies in a survey, would not be
possible on real data because the shear is not constant\footnote{It is
  an open question whether stacking over small areas, in which the
  shear is approximately constant is feasible, although no such
  attempt was made on the GREAT10 data.}; furthermore
in real data the PSF is not constant across images. In addition to these realisations
it was clear that the metrics used to gauge the performance of methods
needed to be more directly related to the quantity of interest when
using weak lensing for dark energy measurements, and that a realistic
spatially varying field would enable full correlations with PSF
quantities (ellipticity and size) to be made (as can be done in real
data). 

To this end GREAT10 introduced the concept of a variable shear
simulation where both the shear field and the PSF varied spatially
across the field of view in a realistic manner. This enabled a variety
of new metrics including a new quality factor that relates the
measured shear \emph{power spectrum} to the true power spectrum 
\be 
Q_{10}=1000\frac{5\times 10^{-6}}{\int {\rm d}\ln \ell |\widetilde
  C^{EE}_{\ell}-C^{EE,\gamma\gamma}_{\ell}|\ell^2}
\ee
in this case the numerator has a well defined meaning as the value
of the denominator that a shape measurement method would need to 
measure the dark energy equation of state
parameter $w_0$ (Linder, 2003) in an unbiased way. In addition the
variable shear field still allows for the constant-shear $m$, $c$ and $q$
parameters to be extracted (one-point estimators of shear as opposed
to spatially variable ones) and some additional metrics defined in
Kitching et al. (2012). The full results of GREAT10 are in Kitching et al. (2012). 

\subsection{Other Public Challenges}
There were several other challenges that were not published but have
been public in the time since STEP1 to the publication of this article
(c. 2012). There have been several incarnations of STEP beyond STEP1 
and STEP2\footnote{\tt http://goo.gl/SVWQ6}. 

STEP1 and STEP2 simulated data as they would appear from a
ground-based telescope since most weak lensing data at the time (and
still now c. 2012) came from ground-based telescopes.  However, significant
effort was also going into weak lensing surveys with the Hubble Space
Telescope (e.g. Schrabback et 
al., 2007; Heymans et al., 2005; Massey et al., 2007).  At the end of STEP2, it was
decided that a similar exercise should be done to obtain a snapshot of
the status of the field of weak lensing shape measurement as it
pertained to space-based data. Space-data is of significantly higher
resolution than ground based data and thus presented a unique set of
both challenges and  advantages. SpaceSTEP (or STEP3), as it was called,
followed nearly the same model as STEP2. The three groups who were
most active in publishing weak lensing results with space based data
all participated. Their methods were shown to be sufficiently
accurate for the size of the surveys at the time; the SpaceSTEP
results were quite similar to the results of STEP2, and thus a
separate paper was never published. 

STEP4 was very similar to GREAT08 in
that simple galaxy models were arranged on a grid, in fact the GREAT08
image simulation code was a conversion of that used for
STEP4. Mirror-STEP was a smaller project designed to test how the mirror size of a
telescope affected shape measurement, and Data-STEP was a link for
people to download and analyses existing weak lensing data. 
In the period between GREAT08 and GREAT10 there was a new
realisation of GREAT08 made `GREAT08 reloaded'. 
\\

\noindent This concludes the short review of previous shape measurement
challenges. We will now present results from the Mapping Dark Matter
competition in the remaining sections. 

\section{Mapping Dark Matter}
\label{Mapping Dark Matter}
The aim of the Mapping Dark Matter competition was to shift the focus
of shape measurement challenges away from verification of methods on
a large amount of realistic data to that of \emph{idea generation}. 
It was run as a competition in partnership with
Kaggle\footnote{{\tt http://www.kaggle.com}} for 2 months from June 
2011 to August 2011.  

The emphasis on idea generation was conceived as a new focus for a number of reasons:
the participation rate of previous challenges was low (of order $10$
to $15$ teams in total) and the methods tried by new teams were either not
directly usable on real data or based on existing methods (the winner of GREAT08 used a method that
was based on a method already published in Kuijken, 1999). The philosophy posed
by the challenge was not as a question to investigate where methods behaved well or
poorly, and not to investigate whether methods can perform on current
or future data, or in particular regime. The goal was to open up the problem to as wide a
community as possible and to encourage open experimentation of ideas,
\begin{center}
\emph{To make shape measurement approachable enough that
  experimentation is easy.}
\end{center}
If it becomes easy to experiment, with useful feedback with minimal
investment in time then new ideas, which previously may
have been difficult to assess due to barriers of entry, become
manageable to try. 

In formulating the challenge a number of guiding principles were
followed, based on the previous challenges (STEP1, STEP2, GREAT08, GREAT10)
\begin{enumerate}
\item 
There must absolutely be no jargon; no FITS images, no `functional
forms', one must not need to know what a star or galaxy is or even why
this measurement is required.
\item 
The simulation must be small enough to download anywhere in the world
over the slowest plausible connection; it should be storable on a USB
stick and accessible via a modem.
\item 
The prize must be desirable; in GREAT08 and GREAT10 the prize was a
piece of hardware (i.e. a laptop or similar), however something more unique may be
more motivating (e.g. a visit to university or attendance of a
conference)\footnote{Although there is a strong correlation between
  those challenges with the most participants and the monetary reward
  for success, in science we can offer something unique: 
  the chance to contribute to our endeavour to understand the
  Universe.}.  
\item 
The question posed to participants, and the data asked for
submissions, must be as simple as possible.
\item 
Given the submission data the metric must be readable and understandable with
no specialist knowledge or jargon. 
\item 
There should be minimal limitation on submission rate. 
\item 
There should be training data that enable participants to test their
methods before submission. 
\item 
The challenge must be blind (participants only use the data made
available to them)\footnote{This is less important for problems where
  the ground-truth is known \emph{a priori} and the task is to develop algorithms to
  recover this in the most efficient manner. But in science domains where
  the ground truth is not known the risk is that algorithms are
  trained to recover simulated input signals only.}. The training data allows for testing in a
controlled way, however if the simulation code is available during a
competition then
arbitrarily large training sets may be generated which would render
results questionable. 
\end{enumerate}
Working under these principles, in partnership with
Kaggle the challenge was formulated as described below. 

\subsection{Description of the simulations}
\label{Description of the simulations}
The Mapping Dark Matter challenge was similar to STEP1 in that it uses
a small number of constant shear images, and simple galaxies models. 
The simulation data were composed of $100$,$000$ simulated galaxies,
each galaxy was presented on a separate PNG postage stamp that was
$48$x$48$ pixels in size. For every galaxy postage stamp there was a
corresponding postage stamp that contained a pixelated representation
of the PSF (a `star' image). 

The $100$,$000$ postage stamps comprised of three groups these were 
\begin{itemize}
\item 
{\bf Training Data}: $40$,$000$ galaxies, these had zero additional
shear and participants were provided with the input ellipticities.
\item
{\bf Public Test Data}: $20$,$000$ galaxies, these had zero additional
shear and participants were ranked in the live leaderboard according
to their score on this data alone.
\item
{\bf Private Test Data}: $40$,$000$ galaxies, these had an additional
shear of $\gamma_1=0.01$ and $\gamma_2=0.01$, participants were not
ranked in the live leaderboard according to their score on this data.
\end{itemize}
To reduce the shape noise contribution to
ellipticity estimates we used the 90 degree rotation transformation as
used in STEP2 such that for every galaxy there was a corresponding
partner that had the same shear but a 90 degree rotated intrinsic
ellipticity in each of the groups. 
\begin{figure*}
  \centerline{\includegraphics[width=1\columnwidth,angle=-90,clip=]{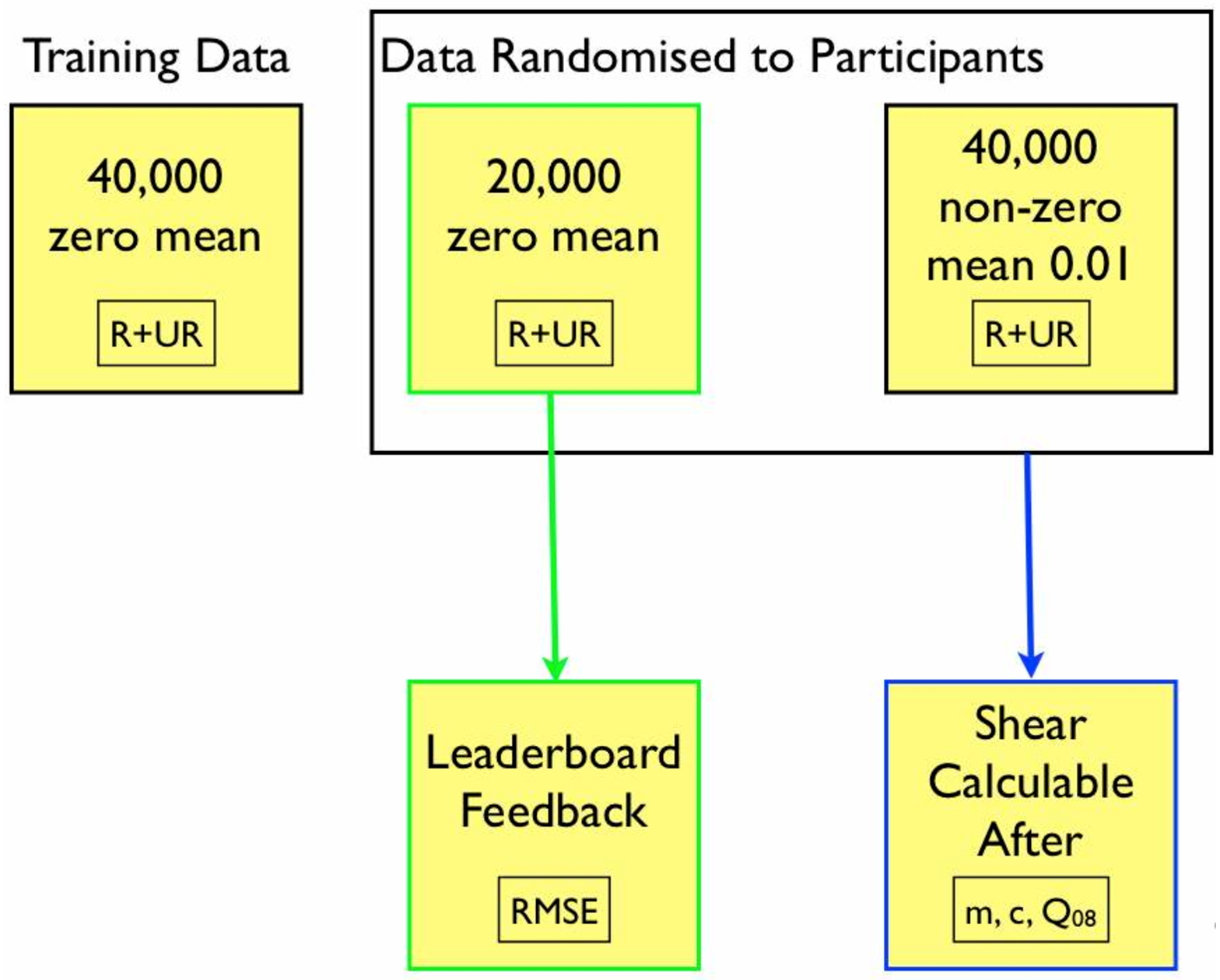}}
  \caption{The simulation structure of Mapping Dark Matter. $R$ and $UR$
    refer to the rotated and unrotated galaxy pairs respectively. The
    $60$,$000$ galaxies in the test data were randomised to participants
    however the leaderboard feedback was provided only on the zero-shear
    group. The leaderboard provided feedback through the RMS error of the
    ellipticity, and the total test data (including the sheared group)
    allows for the constant shear metrics $m$, $c$ and $Q_{08}$ to be
    analysed after the challenge in this article.}
 \label{mdm_structure}
\end{figure*}

The Test Data was randomised so that participants downloaded a set of
$60$,$000$ galaxies and were asked to upload results for all these
galaxies, they were informed that the score was based on $30\%$ of the
data. Participants were asked to provide a CSV file that contained
$60$,$000$ rows where the challenge was, for each galaxy to measure
the ellipticity as accurately as possible. The ellipticity was 
parameterised by $e_1$ and $e_2$, defined as 
\ba 
e_1=\frac{a-b}{a+b}\cos(2\theta)\nn
e_2=\frac{a-b}{a+b}\sin(2\theta)
\ea
where $a$ and $b$ are the semimajor and semiminor axes of the ellipse
and $\theta$ is the position angle. A definition of ellipticity
defined in terms of quadrupole moments was also provided. 
Participants were scored during the challenge using the root mean
squared error between the submitted ellipticity and the true ellipticity 
\be
{\rm RMSE}=\langle (e^{\rm submitted}-e^{\rm true})^2\rangle^{1/2}
\ee
where the average was over all galaxies with zero shear. 
This metric was a measure of methods ability to measure the
ellipticities of galaxies (without recourse to shear), which is the
first order requirement for a good shape measurement method even
though it does not equate to the quantity of interest (the
shear). This metric was also readily understandable, and the
public/private split of the data allows meaningful scores to be
returned on data without shear, whilst at the same time enabling an
investigation into shear after the challenge. 
In Figure \ref{mdm_structure} we show a schematic of the simulation
structure. 

The simulated galaxies were bulge and disk models using the same intensity
profiles presented in the GREAT10 Galaxy Challenge article (Kitching
et al. 2012). The PSF was different for every object where 
the distribution of simulated PSF sizes and
ellipticities were taken from the Jarvis, Schecter
\& Jain (2008) model as described in (Kitching
et al., 2012). We summarise the galaxy and PSF properties in Table \ref{simprop}.
\begin{table}\footnotesize
\centering
\begin{tabular}{|l|c|}
\hline
{\bf Galaxy Property}&{\bf Value}\\
\hline
Postage Stamp Size&$48$x$48$ pixels\\
Signal-to-Noise Ratio&$40$\\
Disk Scale Radius&$4.8$ pixels\\
Ellipticity&$[0.0$, $0.6]$ in $e_1$ and $e_2$\\
\hline
\hline
{\bf Star Property}&{\bf Value}\\
\hline
Moffat $\beta$&$3$\\
FWHM&$[3,4]$ pixels\\
Ellipticity&$[0.01$, $0.1]$ in $e_1$ and $e_2$\\
\hline
\end{tabular}
\caption{A summary of the main parameters that defined the Galaxy and
  PSF models in Mapping Dark Matter. The Galaxy ellipticity
  distribution used was the same as for the GREAT10 Galaxy challenge
  (equation 47 in Kitching et al. 2012), the PSF ellipticities and
  sizes were sampled from the Jarvis Schecter and Jain model as in 
  the GREAT10 Galaxy challenge. The signal-to-noise was
  scaled to match the default SExtractor (Bertin \& Arnouts 1996) 
  {\tt flux\_auto/flux\_err\_auto} parameter combination.}
\label{simprop}
\end{table}

\subsection{Shape measurement results}
\label{Shape measurement results}
The team DeepZot (authors Kirkby and Margala) won the challenge by using a mixture of maximum
likelihood fitting of simple models with a neural net training method
on the ellipticity values (see Appendix A). We provide the data used
to create the results in this Section here {\tt http://great.roe.ac.uk/data/mdm\_figures/}. 

\begin{figure*}
  \includegraphics[width=2.0\columnwidth,clip=]{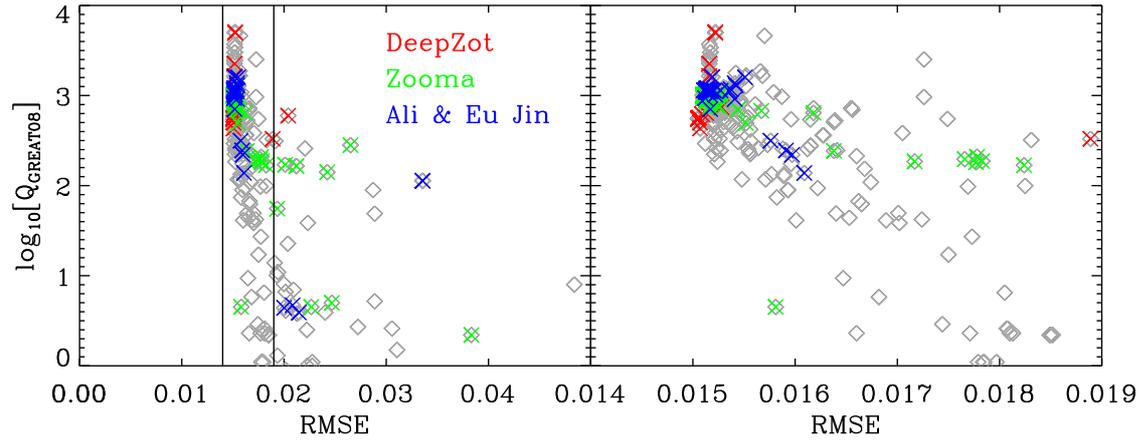}
  \caption{The RMSE and the quality factor $Q_{08}$ for each of the submissions for the top 
    $15$ teams (gray points). We highlight the top 3 teams using red, green and blue points. The 
    right hand panel is a copy of the left hand panel except with an expanded x-axis scale (the region 
    is denoted by the vertical lines in the left hand panel).}
 \label{s2}
\end{figure*}

In Figure \ref{s2} we show the RMSE values for each of the top 15 team's submissions, 
and highlight the top 3 team's submissions. Comparison of the RMSE with the 
quality factor $Q_{08}$ shows a correlation, with a minimum RMSE limited by the 
signal-to-noise of the simulations. The best methods achieve a
$Q_{08}\simeq 5000$; this is a factor of $2$ to $3$ times the highest
quality factor achieved by methods on constant shear simulations before
this challenge: the best reported values published after the GREAT08
challenge to this article are $Q_{08}\simeq 3000$ from
Bernstein (2010) and $Q_{08}\simeq 1300$ from Gruen et al. (2010). 
The best methods achieve RMSE$\simeq 0.015$ this can also 
be compared to the benchmark we used SExtractor (Bertin \& Arnouts,
1996), the source detection and shape measurement technique most
widely used in astronomy, that achieved RMSE$\simeq 0.086$. 

\begin{figure*}
  \includegraphics[width=\columnwidth,clip=]{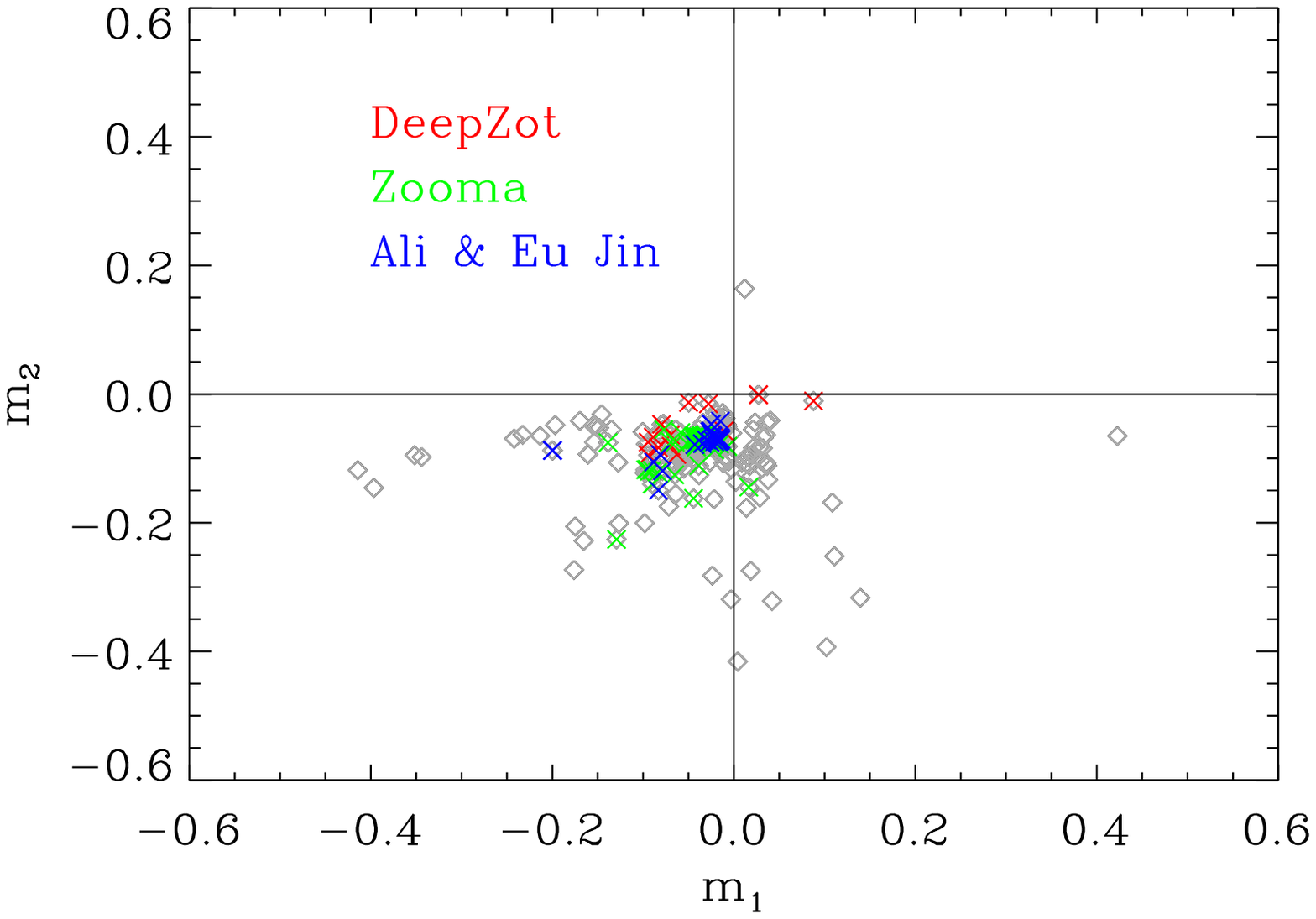}
  \includegraphics[width=\columnwidth,clip=]{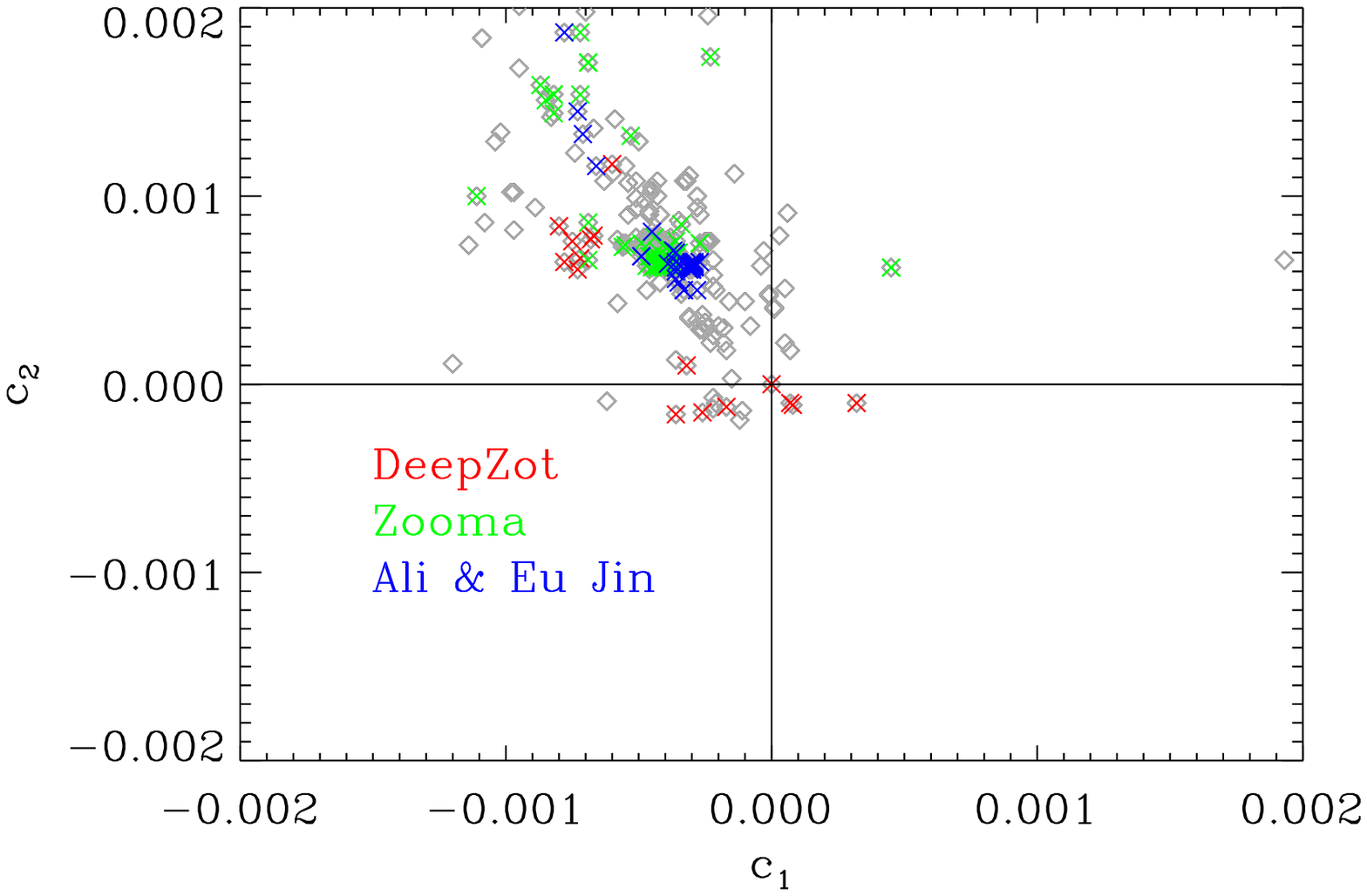}
  \caption{The STEP $m$ and $c$ values for $\gamma_1$ and $\gamma_2$ for each
    submission from the top 15 teams (gray points), we highlight the
    top 3 team's submissions (red, green and blue points).}
 \label{mm}
\end{figure*}

\begin{figure*}
  \includegraphics[width=2\columnwidth,clip=]{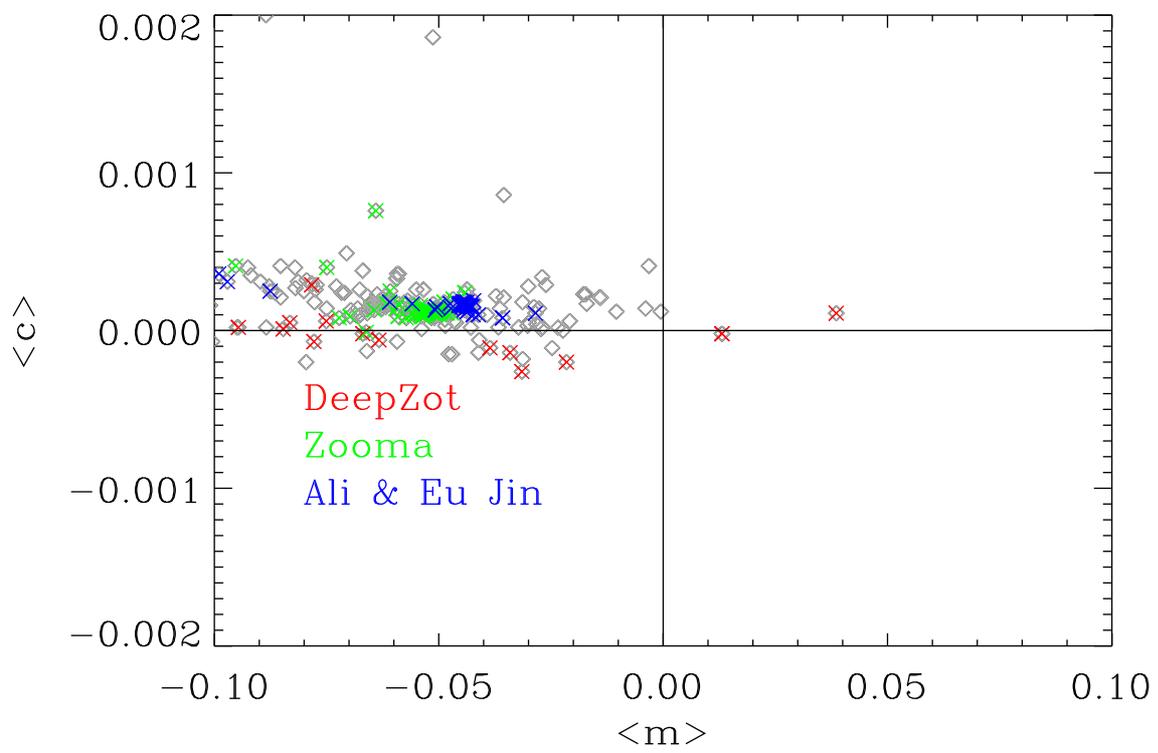}
  \caption{The mean STEP $m$ and $c$ values, averaged over $\gamma_1$ and $\gamma_2$; we 
  show these values for the top 15 team's submissions (gray points) and highlight the top 3 team's submission 
  (red, green and blue points).}
 \label{qm1}
\end{figure*}

\begin{figure*}
  \includegraphics[width=2\columnwidth,clip=]{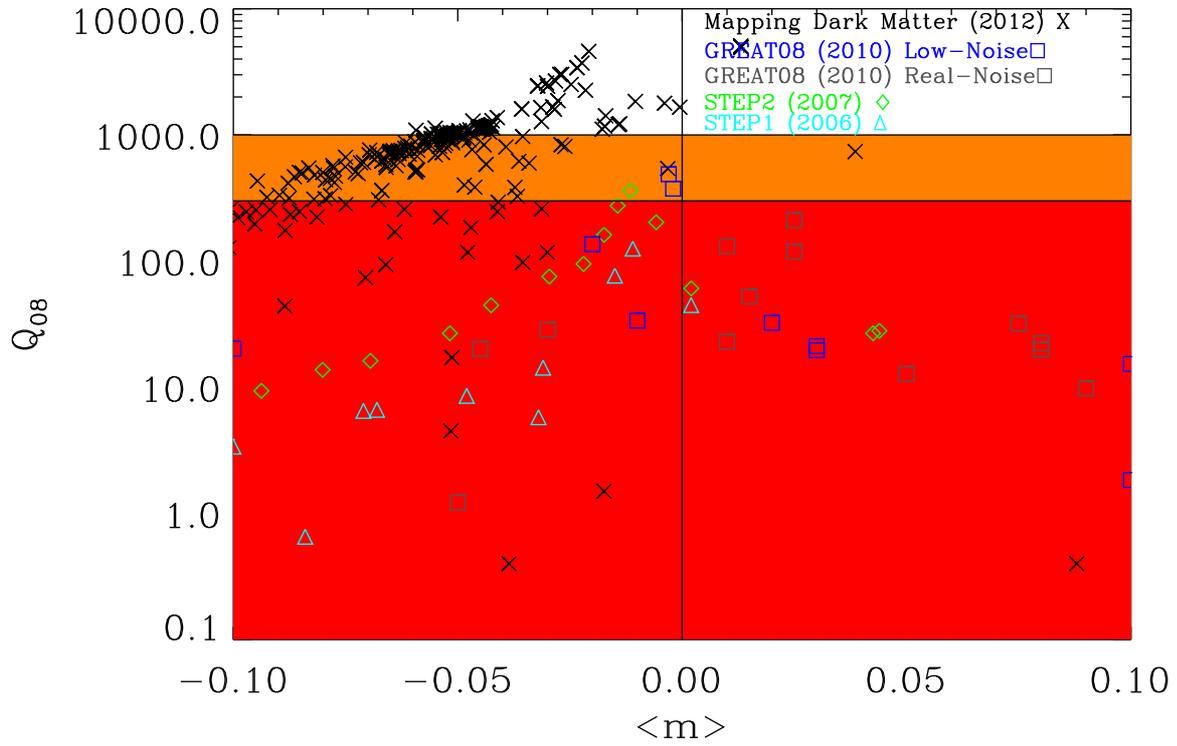}
  \caption{The quality factor $Q_{08}$ and the mean STEP $m$ value; 
  we show this for each submission from the top 15 teams, and also
  show the values from STEP1, STEP2, GREAT08 (`low noise') and GREAT08
  (`real noise') for comparison. This compares each constant shear simulation to
  date. The shaded regions indicate $Q\leq 300$, $300< Q\leq 1000$
  and $Q> 1000$ to help guide the reader. GREAT08 low noise was S/N$=100$ and 
  GREAT08 real noise was S/N$=10$ (using definitions consistent
  with STEP1/STEP2 and Mapping Dark Matter), STEP1
  and STEP2 had S/N$\simeq 10$-$20$, MDM had S/N$=40$.}
 \label{qm2}
\end{figure*}

The RMSE and $Q_{08}$ results are reflected in the STEP parameter results. 
In Figures \ref{mm} and \ref{qm1} we show the STEP $m$ and $c$ values for the top 15 teams, 
and highlight the entries submitted by the top 3 teams; and in Figure \ref{qm2} we 
show how the mean $m$ relates the the quality factor
$Q_{08}$\footnote{We calculate the STEP $Q_{08}$ values using
  $Q_{08}=10^{-4}/\langle(m\gamma^T+c)^2\rangle\simeq 10^{-4}/(\langle
  m^2\gamma^{T,2}\rangle+\langle
  c^2\rangle)=10^{-4}/((\langle
  m\rangle^2+\sigma_m^2)(\langle\gamma^{T}\rangle^2+\sigma_{\gamma}^2)+
  \langle c\rangle^2+\sigma_c^2)$. For STEP1 we have 
  $(\langle\gamma^{T}\rangle,\sigma^2_{\gamma})\simeq (0.033,0.0018)$
  and we have $m$, $\sigma_m$ and $\sigma_c$ values available from
  Heymans et al. (2006), and for
  STEP2 where the shears were sampled from a flat PDF with shears less than
  $|\gamma|<6\%$ we have
  $(\langle\gamma^T\rangle,\sigma^2_{\gamma})\simeq (0.0, 0.00108)$
  and we have $m$, $\sigma_m$, $c$ and $\sigma_c$ values available from 
  Massey et al. (2007); but note that these are only 
  approximations (and there is a term $2\langle mc\gamma\rangle$ in
  the denominator that is zero for STEP1 and STEP2). For GREAT08 the
  values are from Table 4 and Figure C1 ($R_{gp}/R_p=1.4$) of Bridle
  et al., (2010) for `low noise' and Table 5 and Figure C3
  ($R_{gp}/R_p=1.4$) for `real noise'.}. We find that the $c_1$ and 
$c_2$ biases are approximately anti-correlated for most methods, which leads to a partial 
cancellation when showing the average $\langle c\rangle$. The majority
of methods have negative $m_1$ and $m_2$ as well as negative $c_1$ and
$c_2$. We find a general correlation between 
$Q_{08}$ and $\langle m\rangle$, methods that have a small bias also tend to have a high quality factor 
but note that $Q_{08}$ is mainly sensitive to $c$ only.  

\begin{figure*}
  \includegraphics[width=2\columnwidth,clip=]{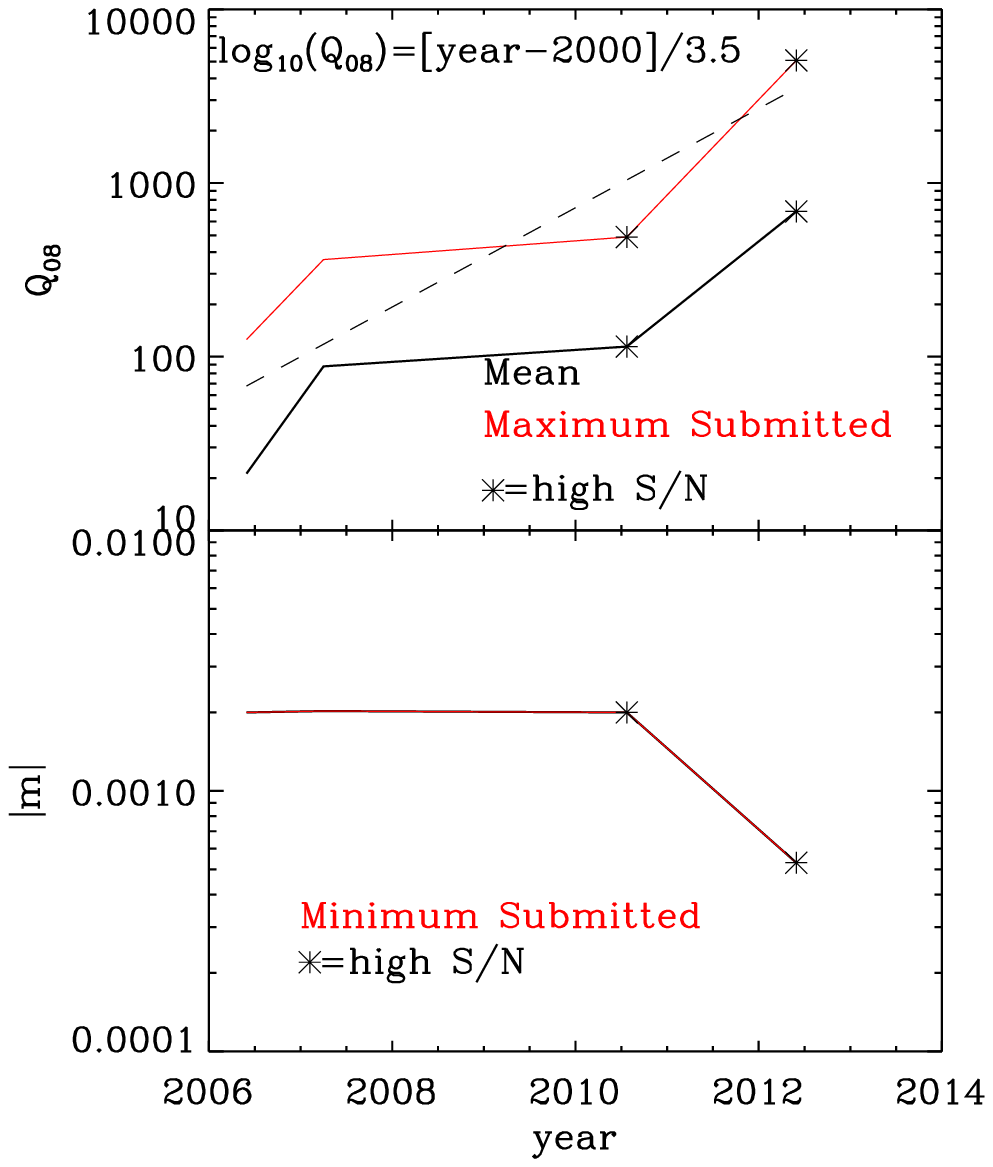}
  \caption{The quality factor $Q_{08}$ and absolute value of the bias 
    $m$ for all \emph{constant shear
    simulations}, STEP1, STEP2, GREAT08 (low noise) and MDM as a function of
    publication date (for variable shear results see Kitching et al.,
    2012). We show the maximum value and the mean value of $Q_{08}$
    and the minimum value of $m$
    over all participants. We show a rule of thumb fit for the
    progression of $Q_{08}$. High signal to noise simulations $\gs 40$ are
    labelled with an asterisk, GREAT08 low noise was S/N$=100$ (using
    a definition consistent with STEP1/STEP2 and Mapping Dark Matter), STEP1
    and STEP2 had S/N$\simeq 10$-$20$, MDM had S/N$=40$.}
 \label{qm3}
\end{figure*}

In Figure \ref{qm3} we show the progression of the $Q_{08}$ and $m$
parameters as a function of time for constant shear simulations (publication
dates for STEP1, STEP2 and GREAT08 were May 2006, March 2007 and July
2010 respectively). We find that
since the year 2000 methods have improved in
accuracy by approximately a factor 10 every approximately 3.5 years
\be 
\log_{10}(Q_{08})\approx \frac{[{\rm year}-2000]}{3.5}.
\ee 
This is similar to, but slightly shallower than, Moore's Law in
computing\footnote{We observe that the timescale for improvement is
  approximately the length of a typical postdoctoral contract (c.2012).}. 

\subsection{Methods}
\label{Methods}
In Appendix A we describe several of the methods submitted to the Mapping Dark Matter challenge. These
innovate over shape measurement methods implemented before this
challenge in a number of ways that we summarise here: 
\begin{enumerate}
\item 
There is extensive use of training methods, in particular neural
networks and Gaussian processes
\item 
There is use of `direct' principal component analysis (PCA) on the data;
extracting the model or vectors from the 
data rather than \emph{a priori} choosing a model
\item 
The use of standard `off the shelf' statistical tools from statistics and particle physics
\end{enumerate}
Most methods employed some variety of model fitting using combinations
of Sersic functions or Gaussian functions and used maximum  
likelihood methods to find best fit parameter combinations. Other
approaches included an implementation of Spergel (2010)  
(submissions by Sogo) and use of wavelets and curvelets (submissions
by Larbi). In Appendix A we describe several methods, 
we refer to methods by the name of the team in the leaderboard (see Figure \ref{leaderboard}). 
We refer to future investigations where the individual tunable aspects
of these algorithms will be tested. 

\subsection{Astrocrowdsourcing}
\label{Astrocrowdsourcing}
\begin{figure*}
  \centerline{\includegraphics[width=2\columnwidth,angle=-90,clip=]{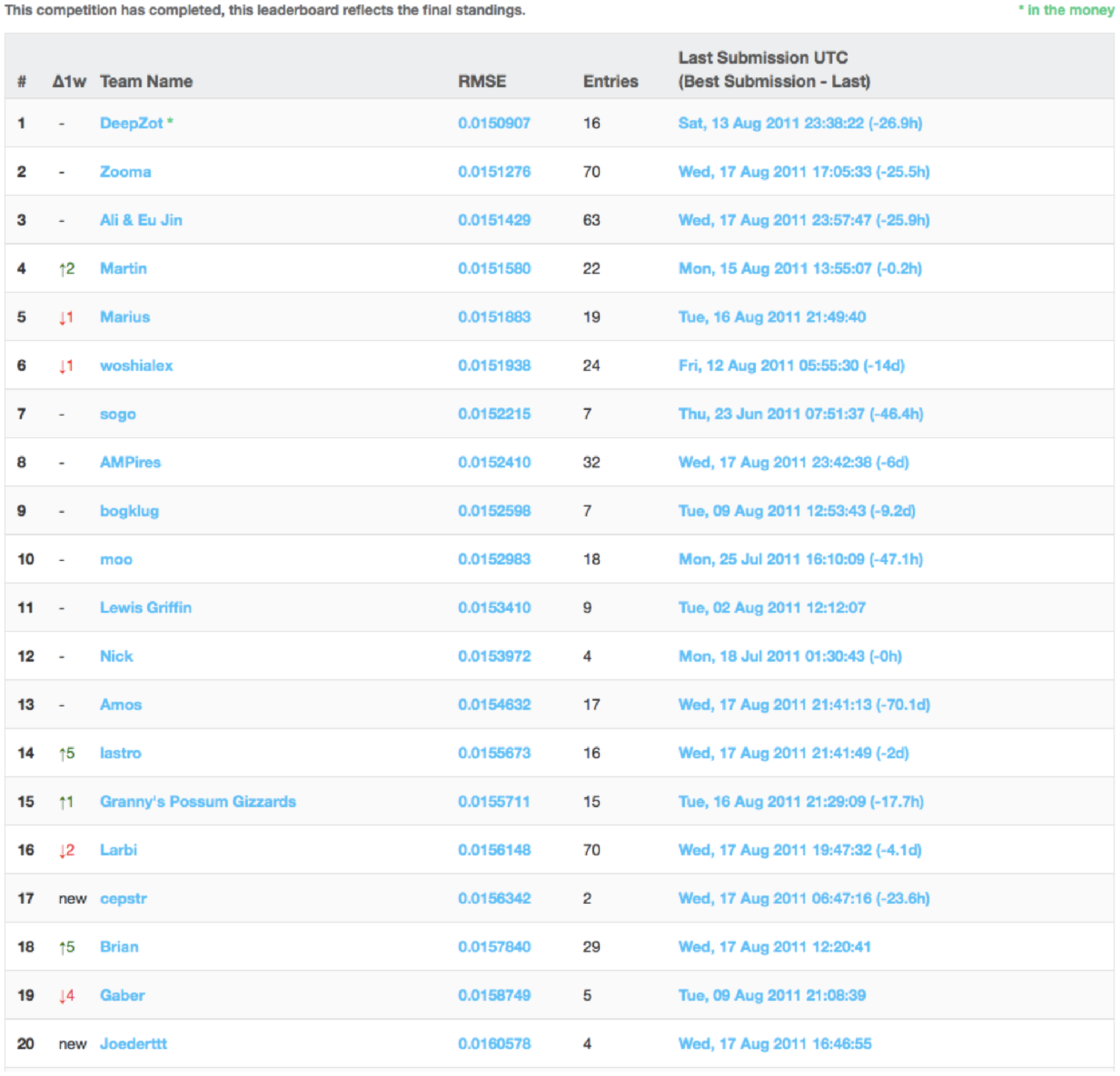}}
  \caption{The leaderboard at the end of the Mapping Dark Matter challenge, from {\tt http://www.kaggle.com/c/mdm}.}
 \label{leaderboard}
\end{figure*}

In Figure \ref{leaderboard} we show the leaderboard at the end of the challenge period. We 
highlight the number of submissions, as well as the competitive submission 
behaviour of the participants which is evident in the submission dates and times. 
The majority of the participants were not experienced in astronomy or cosmology, this marks a 
major change in the demonstrated accessibility of weak lensing data
analysis and is a successful  
example of crowdsourcing astronomical algorithm development what we refer to as `astrocrowdsourcing'. 

\begin{figure*}
  \centerline{\includegraphics[width=1.5\columnwidth,angle=-90,clip=]{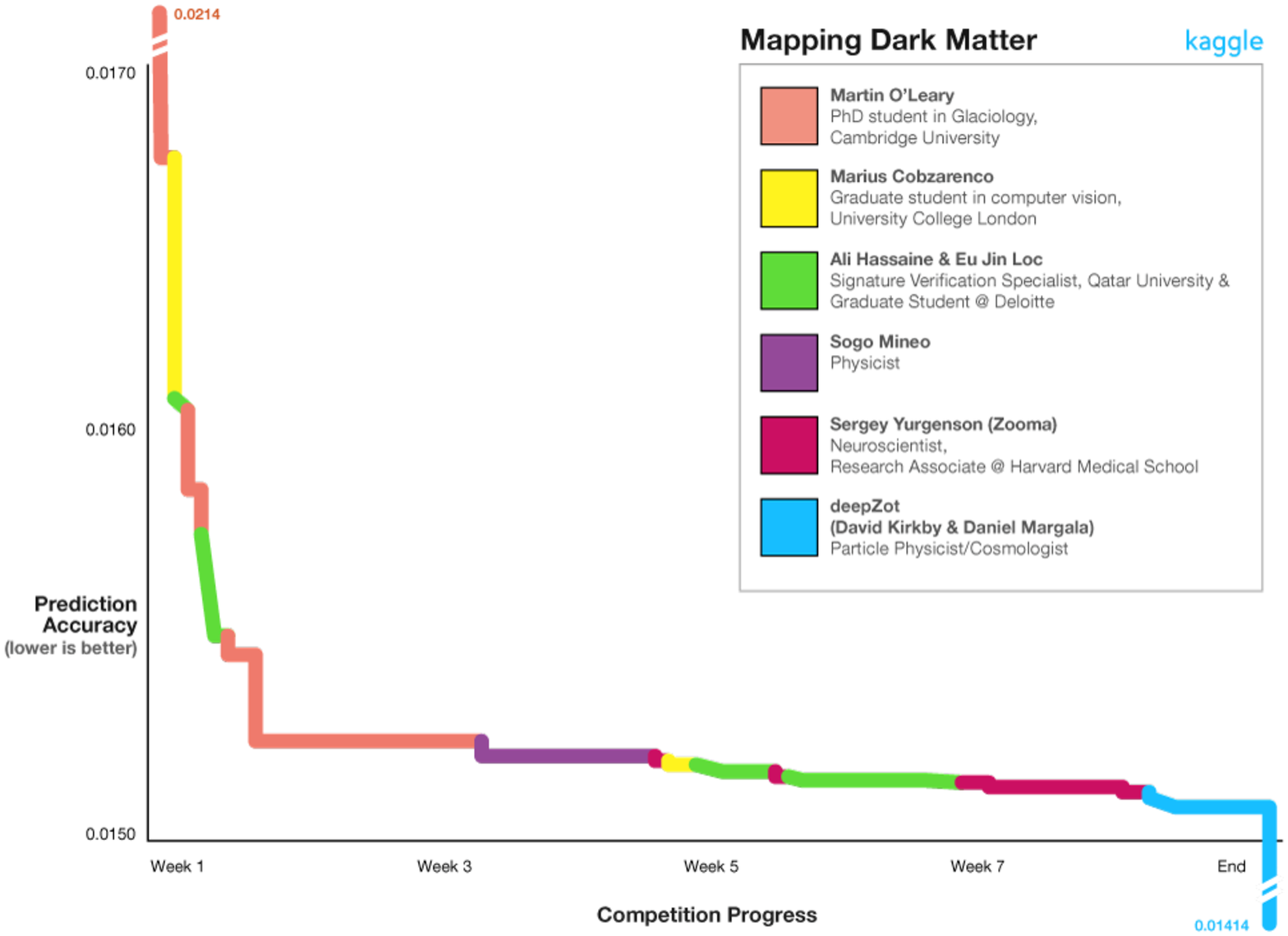}}
  \caption{The change in the best score as a function of time during
    the Mapping Dark Matter challenge.}
 \label{CumuK}
\end{figure*}

\begin{figure*}
  {\includegraphics[width=0.7\columnwidth,angle=90,clip=]{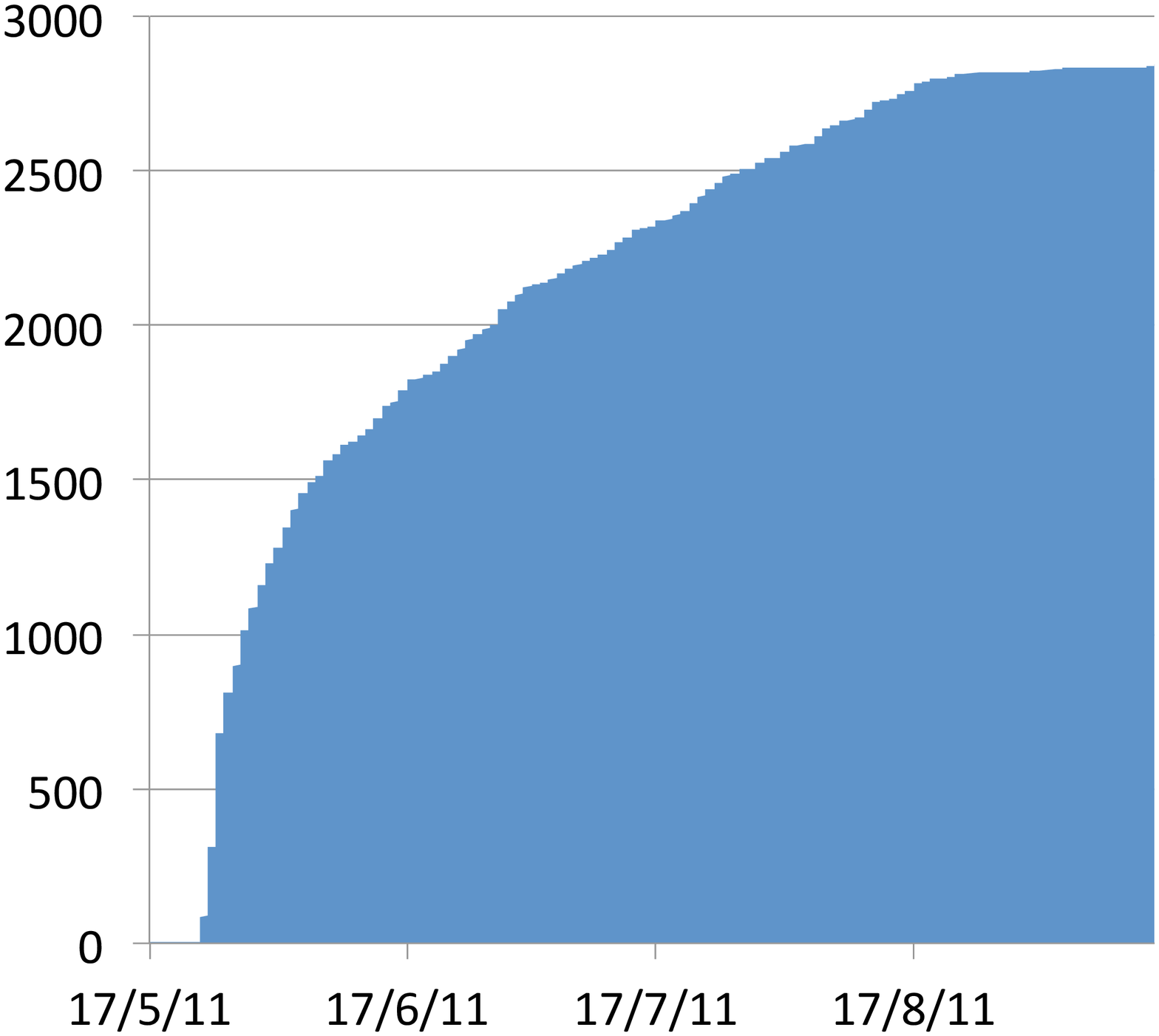}}
  {\includegraphics[width=0.7\columnwidth,angle=90,clip=]{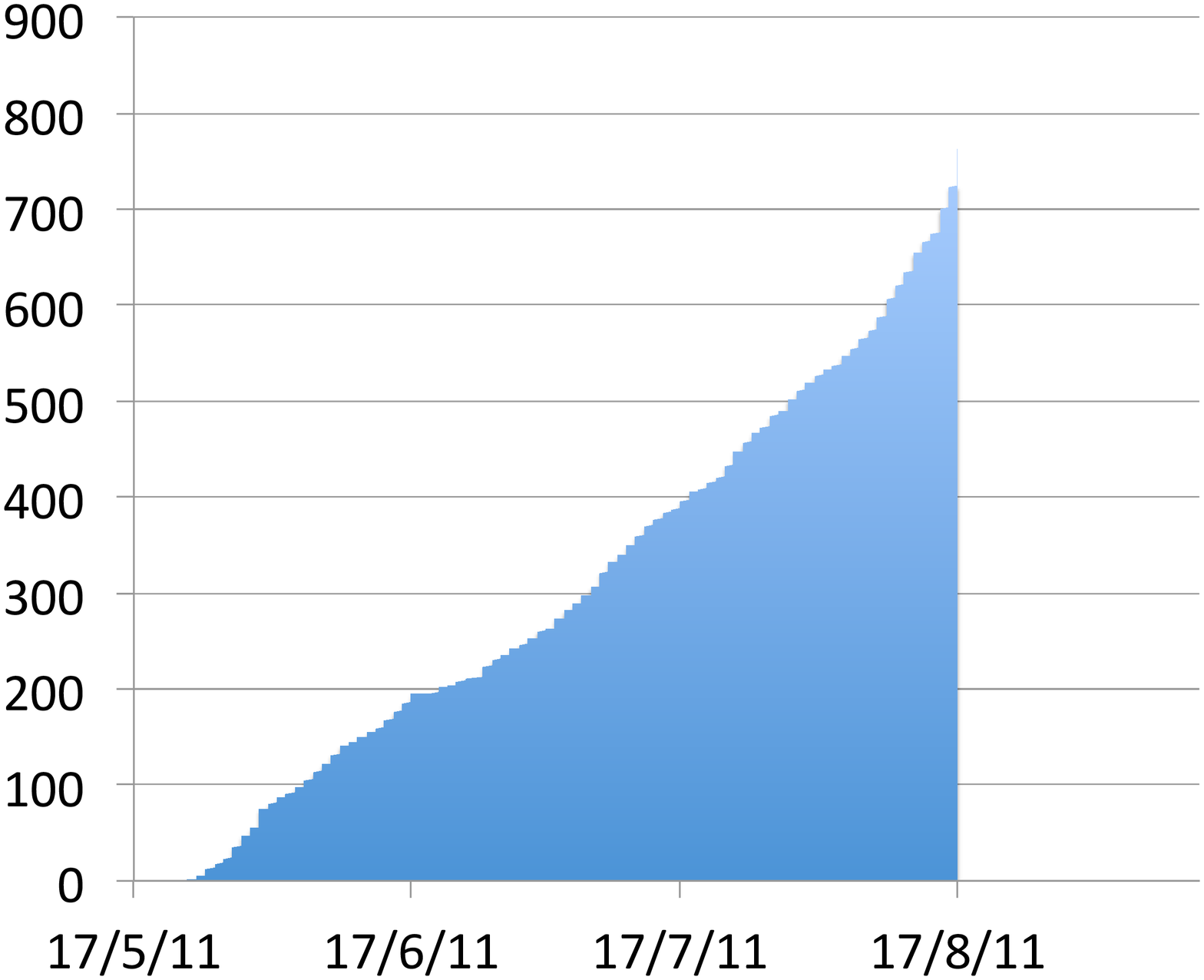}}
  \caption{The right hand panel shows the cumulative number of downloads of the data as a function of time
    over the Mapping Dark Matter challenge period and beyond, the left hand panel shows the cumulative number of submissions as a function
    of time over the same period, the sharp cut-off is when the challenge ended.}
 \label{DownSubs}
\end{figure*}

In Figure \ref{CumuK} we show how the top (best) score changed as a function of 
time and highlight which participant had submitted this score at each moment in the challenge. 
We highlight from this Figure several aspects, that are symptomatic of a challenge that 
has been successfully built to engage participants 
\begin{itemize}
\item 
Rapid improvement early in the challenge. In two weeks the score rapidly improved, and the fractional change was the 
most significant. This reflects participants that apply existing
methodology, and have been engaged early 
\item 
The `Roger Bannister Effect'. Where imaginary barriers are broken by
one team that motivates others to also achieve the same (in analogy to
the `impossible' 4 minute mile that once achieved by Roger Bannister
was subsequently achieved by several others in a short span of time
and by over 1000 others to this date). This is seen in the
period in weeks 1 to 3 when Martin O'Leary held the lead for sometime
after which a succession of lead-changes were seen.
\item 
Alternating/battling teams. We see the lead change hands between two
or several teams alternately. 
\end{itemize} 
Similarly demonstrative of the accessibility of the Mapping Dark
Matter challenge is the download rate of the data over the challenge and the submission 
rate from participants shown in Figure \ref{DownSubs}. The participation rate was constant over the challenge with approximately 
$13$ submissions per day over the $2$ month period. The data download followed a different trend where in the first week $1500$ 
downloads were made, which then reached an equilibrium of approximately $26$ downloads per day.

\section{Conclusions}
\label{Conclusions}
In this paper we present a review of weak lensing shape measurement
challenges to date, including the Shear TEsting Programmes (STEP1 and
STEP2) and the GRavitational lEnsing Accuracy Testing competitions
(GREAT08 and GREAT10). From 2006 we have seen a change in emphasis
from competitions that test methods on fully realistic
images to creating simulations that provide simple development environments
for methods. We also present results from the Mapping Dark Matter
competition, which by simplifying the shape measurement challenge to
the point where it was accessible to a wide audience, 
generated new avenues of investigation for shape
measurement by attracting over 700 submissions over 2 months and saw a
factor of $3$ improvement in shape measurement accuracy on high
signal-to-noise galaxies, over previously published results, and a
factor $10$ improvement over methods tested on blind simulations.

\section*{Acknowledgements}
We thank Kaggle for contributing to all aspects of the Mapping Dark
Matter Challenge, in particular Anthony Goldbloom and Jeremy Howard. 
We thank all participants of the challenge and all those
that downloaded the data. TDK is supported by a Royal Society
University Research Fellowship, and was supported by an Royal
Astronomical Society 2010 Fellowship for the majority of this
work. RM is supported by a Royal Society
University Research Fellowship. The Mapping Dark Matter workshop in partnership with the 
GREAT10 challenge workshop was funded by JPL, 
run under a contract for NASA by Caltech, and hosted at IPAC at
Caltech. We thank Harry Teplitz and Helene Seibly for local organisation of
the workshop. We thank Sree Balan and Konrad Kuijken for development
of the GREAT08 code that was used to generate the data. We thank Alan
Heavens, Andy Taylor, Mandeep Gill, Barney Rowe, Sarah Bridle for useful
discussions. 

%% References


\bibliographystyle{model1-num-names}
\bibliography{<your-bib-database>}

\begin{thebibliography}{00}

\bibitem[Albrecht et al.(2006)]{2006astro.ph..9591A} Albrecht, A., 
Bernstein, G., Cahn, R., et al.\ 2006, arXiv:astro-ph/0609591 

\bibitem[Peacock et al.(2006)]{2006ewg3.rept.....P} Peacock, J.~A., 
Schneider, P., Efstathiou, G., et al.\ 2006, ''ESA-ESO Working Group on 
''Fundamental Cosmology'', Edited by J.A.~Peacock et al.~ ESA, 2006.'',  

\bibitem[Massey et al.(2010)]{2010RPPh...73h6901M} Massey, R., Kitching, 
T., \& Richard, J.\ 2010, Reports on Progress in Physics, 73, 086901 

\bibitem[Bartelmann \& Schneider(2001)]{2001PhR...340..291B}
  Bartelmann, M., \& Schneider, P.\ 2001, Phys REP, 340, 291 

\bibitem[Weinberg et al.(2012)]{2012arXiv1201.2434W} Weinberg, D.~H., 
Mortonson, M.~J., Eisenstein, D.~J., et al.\ 2012, arXiv:1201.2434 

\bibitem[Kitching et al.(2010)]{2010arXiv1009.0779K} Kitching et
  al.,2011, Annals of Applied Statistics 2011, Vol. 5, No. 3,
  2231-2263

\bibitem[Kaiser et al.(1995)]{1995ApJ...449..460K} Kaiser, N., Squires, G., 
\& Broadhurst, T.\ 1995, ApJ, 449, 460 

\bibitem[Melchior et al.(2011)]{2011MNRAS.412.1552M} Melchior, P., Viola, 
M., Sch{\"a}fer, B.~M., \& Bartelmann, M.\ 2011, MNRAS, 412, 1552 

\bibitem[Kuijken(1999)]{1999A&A...352..355K} Kuijken, K.\ 1999, AAP, 352, 355 

\bibitem[Refregier(2003)]{2003MNRAS.338...35R} Refregier, A.\ 2003, MNRAS, 
338, 35 

\bibitem[Miller et al.(2007)]{2007MNRAS.382..315M} Miller, L., Kitching, 
T.~D., Heymans, C., Heavens, A.~F., 
\& van Waerbeke, L.\ 2007, MNRAS, 382, 315 

\bibitem[Heymans et al.(2006)]{2006MNRAS.368.1323H} Heymans, C., et al.\ 2006, MNRAS, 368, 1323 

\bibitem[Massey et al.(2007)]{2007MNRAS.376...13M} Massey, R., et al.\ 2007, MNRAS, 376, 13 

\bibitem[Kitching et al.(2012)]{2012arXiv1202.5254K} Kitching, T.~D., 
Balan, S.~T., Bridle, S., et al.\ 2012, arXiv:1202.5254 

\bibitem[Bridle et al.(2009)]{2009AnApS...3....6B} Bridle, S., et al.\ 2009, Annals of Applied Statistics, 3, 6
 
\bibitem[Bridle et al.(2010)]{2010MNRAS.405.2044B} Bridle, S., et al.\ 2010, MNRAS, 405, 2044 

\bibitem[Massey 
\& Refregier(2005)]{2005MNRAS.363..197M} Massey, R., \& Refregier, A.\ 2005, MNRAS, 363, 197 

\bibitem[Ferry et al.(2008)]{2008APh....30...65F} Ferry, M., Rhodes, J., 
Massey, R., et al.\ 2008, Astroparticle Physics, 30, 65 

\bibitem[Dobke et al.(2010)]{2010PASP..122..947D} Dobke, B.~M., Johnston, 
D.~E., Massey, R., et al.\ 2010, PASP, 122, 947 

\bibitem[Kitching et al.(2008)]{2008MNRAS.390..149K} Kitching, T.~D., 
Miller, L., Heymans, C.~E., van Waerbeke, L., 
\& Heavens, A.~F.\ 2008, MNRAS, 390, 149 

\bibitem[Linder(2003)]{2003astro.ph.11403L} Linder, E.~V.\ 2003, 
arXiv:astro-ph/0311403 

\bibitem[Schrabback et 
al.(2007)]{2007A&A...468..823S} Schrabback, T., Erben, T., Simon, P., et al.\ 2007, AAP, 468, 823 

\bibitem[Heymans et al.(2005)]{2005MNRAS.361..160H} Heymans, C., Brown, 
M.~L., Barden, M., et al.\ 2005, MNRAS, 361, 160 

\bibitem[Massey et al.(2007)]{2007ApJS..172..239M} Massey, R., Rhodes, J., 
Leauthaud, A., et al.\ 2007, ApJS, 172, 239 

\bibitem[Jarvis et al.(2008)]{2008arXiv0810.0027J} Jarvis, M., Schechter, 
P., \& Jain, B.\ 2008, arXiv:0810.0027 

\bibitem[Spergel(2010)]{2010ApJS..191...58S} Spergel, D.~N.\ 2010, ApJS, 
191, 58 

\bibitem[22]{1996ApJ} Bertin, E. \& Arnouts, S. 1996: SExtractor:
  Software for source extraction, Astronomy \& Astrophysics Supplement
  317, 393 

\bibitem[Bernstein(2010)]{2010MNRAS.406.2793B} Bernstein, G.~M.\ 2010, 
MNRAS, 406, 2793 

\bibitem[Gruen et al.(2010)]{2010ApJ...720..639G} Gruen, D., Seitz, S., 
Koppenhoefer, J., \& Riffeser, A.\ 2010, ApJ, 720, 639 

\bibitem[Cobzarenco (2012)]{mthesis} Cobzarenco, M., 2012, Masters
  Thesis, UCL, Supervisor: Yee Whye Teh

\end{thebibliography}




\newpage
\newpage

\appendix
\section{Method Descriptions}
In this Section we describe several of the new methods submitted to
the Mapping Dark Matter challenge. We have shortened URLs where needed
for typographical reasons. 

\subsection{Ali \& Eu Jin: A. Hassa\"{i}ne and E. J. Lok}
This method used techniques taken from two fields: 
signature verification/writer identification and soundtrack
restoration, along with other methods specifically developed for the
challenge\footnote{For a list of predictors used see 
{\tt http://goo.gl/GjDXC}, code 
can be accessed from here {\tt http://goo.gl/Ty4UM}.}
A short list of the predictors used were 
\begin{enumerate}
\item 
Computing $e_1$ and $e_2$ for a Gaussian-smoothed thresholded version of galaxy images.
\item 
Computing $e_1$ and $e_2$ for a Gaussian-smoothed thresholded version of star images
\item 
Computing $e_1$ and $e_2$ for a convolved version of galaxy images.
\item 
Creating structuring element from the star images and using it to perform basic morphological operations on the galaxies.
\item 
Computing directions and curvatures of both galaxy and star images.
\item 
Computing chain codes and edges features from both galaxy and star images.
\end{enumerate} 
Several of these predictors are also computed on a $\pi/4$ (45 degree) rotated version of the galaxy images.
Whenever a method has one or more parameters, each possible value of the parameter was used to generate a separate predictor.
Finally all these predictors were combined via linear fit. 

\subsection{woshialex: Q. Liu}
This method uses the idea of reconstructing the galaxy image with a
model including the parameters $e_1$ and $e_2$, and fit the best
parameters. The model is built with physics insights about the shape
of the galaxy, its intensity distribution, and convolution. It starts
from a good initial guess of the parameters by other simple methods (in this case, the unweighted quadrupole moments),  and then generates
a galaxy image based on the model we build (try to reproduce the
image). The parameters of the model are then tuned to minimise the
difference between the generated galaxy image and the original image
with the difference measured by the $\chi^2$. The minimisation is
achieved using the {\tt nlopt} package \footnote{
  http://ab-initio.mit.edu/wiki/index.php/NLopt }. There are more
descriptions on the challenge forum\footnote{http://goo.gl/uZDcL}. The
main steps of the algorithm are as follow, 

\begin{enumerate}
\item Fit the star image using a functional form $1/(1+r^2)^3$ where $r^2=(x-x_{c1})^2/a_1^2+(y-y_{c1})^2/b_1^2$.
\item Generate an initial galaxy image using a functional form
  $\exp({(x-x_{c2})^2/a_2^2+(y-y_{c2})^2/b_2^2})^{1/2}$, with initial
  parameters from the quadrupole moments. 
\item Convolve this initial galaxy image with the fitted star
  function, and we obtain a new galaxy image. Our goal is to reproduce
  the provided galaxy image by tuning the parameters in our model.  
\item Use the nonlinear optimisation method provided in the package
  {\tt nlopt} to minimise $\chi^2$ (the sum of squares of the difference
  between the generated image and the provided image at each pixel) by
  tuning parameters $a_2$ and $b_2$ and others. Then $e_1$ and $e_2$
  are calculated from $a_2$ and $b_2$. 
\end{enumerate}
A neural network training is applied to improve the final fitted
results of $e_1$ and $e_2$, but no improvement is found. So the final
reported results are just the fitted value. 

\subsection{DeepZot: D. Kirkby and D. Margala}
This methods consists of two steps. The first step is a
pixel-level maximum-likelihood fit to each star and galaxy image to
extract shape parameters (including the ellipticities) and their
covariance matrix. The second step is to feed a subset of the fit
outputs into a neural network (configured for regression rather than
classification) that is trained to provided corrections to the fitted
ellipticities. Only the second step was varied to produce different submissions. 

Skipping the second step entirely and using the fitted outputs
directly gave scores of $0.0151432$ (public) and $0.0152543$ (private), so
the fit is doing most of the work in estimating ellipticity, but the
neural net provided a small but
significant improvement (that meant that DeepZot won the challenge). 

The fit minimisation engine (Minuit) and NN engine (TMVA) used are
both available as part of the open source (LGPL) ROOT data analysis
framework ({\tt http://root.cern.ch}) that is widely used by particle
physicists. 

For more details of this method see the GREAT10 results paper Kitching et al. (2012). 

\subsection{Zooma: S. Yurgenson}
This method finds the principal components of the galaxy images, and 
find those eigenfunctions that maximally correlate with the
ellipticities. It can be summarised in the following simple steps 
\begin{enumerate}
\item
First the centers of the galaxies and stars were found using a weighted mean (moments) with a threshold.
Images were then recentered using spline interpolation.
\item
From the image stacks the primary principal components were calculated. 
\item 
The component amplitudes were then entered into a neural net with $e_1$ and $e_2$ as targets.
This was repeated multiple times choosing several different network configurations (using the training data) 
to find the ``best'' networks, by slightly changing centering methods and networks parameters.
\item 
The mean prediction over multiple networks was calculated. The best scoring submission (smallest RMSE) 
was a mean of $35$ predictors, each with RMSE$<0.015$ on the training set. 
\end{enumerate}
For a detailed method description with Matlab code snippets see {\tt http://goo.gl/nLGmG}.

\subsection{Grannys Possum: B. L. Cragin}
This method used a simple model similar to woshialex's. 
The images were additionally de-noised using principal component
analysis (PCA) decomposition and
retaining only the first 16 terms in the eigenfunction expansion,
prior to all other analysis. The star images were then fit 
(using simple $\chi^2$ minimisation with a top-hat weighted to the middle 
half of the image) to an
elliptical Moffat distribution. This gave the semimajor axes $a$, $b$ and 
position angle $\theta$ and hence
ellipticities of the stars. 
This fit to the star image was then convolved with another elliptical
Moffat function (representing the sought-after, pre-convolution
galaxy), the parameters of which were then iterated for best fit of
the result to the observed galaxy. 

Considerable improvement was obtained in the form of a ``three-epsilon
model''. In this model an elliptical Moffat profile was also fit to the
observed galaxy (with convolution), yielding a third pair of ellipticity values. 
A simple linear regression and a 
Support Vector Machine (SVM) were then used to predict the pre-convolved ellipticities.
After its kernel and
target parameters were optimised for least-cross-validation error, the
SVM performed as well as linear regression, but did not
outperform it\footnote{The code for this was written in R, with the exception
of the PCA decomposition part which was done in SciLab.}.

\subsection{AMPires: A. M. Pires}
In this investigation several methods were attempted. The best results with smallest 
RMSE were obtained  with a
combination of principal components analysis and multiple linear
regression. The main steps were 
\begin{enumerate}
\item 
A central window on each image (different sizes for stars
and galaxies) was used, and moved in the image plane by one pixel in every direction, 
generating $9$ images for every galaxy and every star. 
\item 
The 9 images were transformed into 9 vectors, and a principal
components analysis was performed in these vectors, the first 3 eigenfunctions 
were then kept. After this
step there are 3 sets of images for the galaxies and 3 sets of images for
the stars. The first set is similar to the result of applying a low
pass filter to the original images, the second and the third may be
interpreted as the result of applying two Sobel filters. 
\item 
A further principal component analysis was then applied, this time to each of
the 6 sets of 40000 images described in 2. 
\item 
The final step was to build a linear regression model using the
first two components from the six sets as explanatory variables and
the ellipticities as response variables. At this stage second order
interactions and powers up to 3 were included. It was also necessary to take into
account the structure of the components to build a sensible model. 
\end{enumerate}
These final steps still have scope for improvement and optimisation. 

\subsection{Marius: M. Cobzarenco}
We experimented with a number of variations
around a generative probabilistic model of PSFs and galaxies. The
method is described at length this Masters thesis Cobzarenco (2011)
{\tt http://goo.gl/woh5s}. The basic model was built around a sum of
S\'ersic profiles with added Gaussian noise. The S\'ersic profiles
were parametrized in terms of $I_0 , k, R$ and $n$  (see for example
the GREAT10 results paper, Kitching et al., 2012) together with
$\sigma^2$ (the variance of the noise) and $\mu(x, y)$ (the
coordinates of the center of the object). I used a conjugate gradient
algorithm to optimize for the maximum of the posterior distribution
(MAP estimates of seven parameters per image). Two of the variations
submitted were: 
\begin{enumerate}
\item \textbf{Individual}: Learning the parameters for the shape of the stars
to reproduce the observed image of the star. Then
learning the parameters for the shape of the galaxy
to reproduce the observed image of the galaxy.
\item \textbf{Joint}: Learning the parameters for the shape of the star
to reproduce the observed image of the star, and
at the same time, learning the parameters for the
shape of the galaxy, such that when convolved with
the star reproduces the image of the galaxy. The
convolution was done numerically.
\end{enumerate}
The final step common to both versions was to fit a Sparse Gaussian
process (Snelson and Ghahramani 2006) to \emph{learn} the mapping
between the MAP parameter estimates and the PSFs/galaxy
elipiticities. 

\subsection{Martin: M. O'Leary}
This method used a linear combination of results from a collection of
disparate approaches. These consisted primarily of maximum-likelihood
estimates of parameters for assumed functional forms. This approach
was motivated by promising early results. 

In the simplest iteration of this technique, both the galaxy and
kernel images were fitted individually using MLE as the sum of
normally distributed white noise and a Gaussian kernel. Noise
parameters were then discarded, and the kernels deconvolved
analytically. Applying a linear correction to the results of this
approach yielded an RMSE of $0.0169$ ($0.0168$ private), indicating that
the approach was viable. This value was reduced to $0.0156$ ($0.0158$
private) by calculating the principal components of both the galaxy
and kernel images, and introducing the first six components from each
as additional variables in the regression. 

Additional contributions to the final `blend' included MLE fits using
both Sersic and De Vaucouleurs profiles for the galaxies, and both
Gaussian and Moffatt profiles for the kernels. Two techniques were
used for deconvolution. In the first, the kernel was fitted initially,
and deconvolution was performed numerically, using the Richardson-Lucy
algorithm. The parameters for the galaxy were then determined from the
deconvolved image. In the second technique, parameters for both the
galaxy and kernel were fitted simultaneously, based on the convolution
of both images. This approach was considerably more computationally
intensive, but provided slightly better results. 

The final blend was computed using linear regression on all solutions,
as well as the principal components previously mentioned. To avoid
overfitting, forward stepwise variable selection was employed, using
the Bayesian Information Criterion. Regression and variable selection
were performed separately for $e_1$ and $e_2$, and results from each
variable were included in regressions for the other. This resulted in
a final RMSE of $0.0150$ ($0.0152$ private). 

\end{document}